# CBTOPE2: An improved method for predicting of conformational B-cell epitopes in an antigen from its primary sequence

Anupma Pandey[1#], Megha[1#], Nishant Kumar[1#], Ruchir Sahni[1,2], Gajendra P. S. Raghava *[1]

1. Department of Computational Biology, Indraprastha Institute of Information Technology, Okhla Phase 3, New Delhi-110020, India.

2. Indian Institute of Science Education and Research (IISER) Pune, Dr Homi Bhabha Road, Pune, Maharashtra, 411008, India.

**Mailing Address of Authors**

Anupma Pandey (AP): anupma23019@iiitd.ac.in    ORCID ID: https://orcid.org/0009-0001-3477-6443

Megha (M): megha23125@iiitd.ac.in    ORCID ID: https://orcid.org/0009-0001-8371-6797

Nishant Kumar (NK): nishantk@iiitd.ac.in    ORCID ID: https://orcid.org/0000-0001-7781-9602

Ruchir Sahni (RS): ruchir.sahni@students.iiserpune.ac.in    ORCID ID: https://orcid.org/0000-0002-9771-5496

Gajendra P. S. Raghava (GPSR): raghava@iiitd.ac.in    ORCID ID: https://orcid.org/0000-0002-8902-2876

**# Equal Contribution**

**\*Corresponding Author**

Prof. Gajendra P. S. Raghava

Head and Professor

Department of Computational Biology

Indraprastha Institute of Information Technology, Delhi

Okhla Industrial Estate, Phase III (Near Govind Puri Metro Station)

New Delhi, India – 110020 Office: A-302 (R&D Block)

Phone: 011-26907444

Email: raghava@iiitd.ac.in

Website: http://webs.iiitd.edu.in/raghava/

# 1. Abstract

In 2009, our group pioneered a novel method CBTOPE for predicting conformational B-cell epitopes in a protein from its amino acid sequence, which received extensive citations from the scientific community. In a recent study, Cia et al. (2023) evaluated the performance of conformational B-cell epitope prediction methods on a well-curated dataset, revealing that most approaches, including CBTOPE, exhibited poor performance. One plausible cause of this diminished performance is that available methods were trained on datasets that are both limited in size and outdated in content. In this study, we present an enhanced version of CBTOPE, trained, tested, and evaluated using the well-curated dataset from Cai et al. (2023). Initially, we developed machine learning-based models using binary profiles, achieving a maximum AUC of 0.58 on the validation dataset. The performance of our method improved significantly from an AUC of 0.58 to 0.63 when incorporating evolutionary information in the form of a Position-Specific Scoring Matrix (PSSM) profile. Furthermore, the performance increased from an AUC of 0.63 to 0.64 when we integrated both the PSSM profile and relative solvent accessibility (RSA). All models were trained, tested, and optimized on the training dataset using five-fold cross-validation. The final performance of our models was assessed using a validation or independent dataset that was not used during hyperparameter optimization. To facilitate scientific community working in the field of subunit vaccine, we develop a standalone software and web server CBTOPE2 (https://webs.iiitd.edu.in/raghava/cbtope2/).

## Author's Biography

1. Anupma Pandey is currently pursuing a Master's degree in Computer Science Engineering at the Department of Computer Science Engineering, Indraprastha Institute of Information Technology, New Delhi, India.
2. Megha is currently pursuing a Master's degree in Computer Science Engineering at the Department of Computer Science Engineering, Indraprastha Institute of Information Technology, New Delhi, India.
3. Nishant Kumar is currently working as Ph.D. in Computational biology from Department of Computational Biology, Indraprastha Institute of Information Technology (IIIT), New Delhi, India.

4. Ruchir Sahni is currently studying as an integrated BS-MS student at Indian Institute of Science Education and Research (IISER) Pune, India. He is currently working as an Intern on Project position at Department of Computational Biology, Indraprastha Institute of Information Technology (IIIT), New Delhi, India.
5. Gajendra P. S. Raghava is currently working as Professor and Head of Department of Computational Biology, Indraprastha Institute of Information Technology (IIIT), New Delhi, India.

## 2. Introduction

The human immune system is composed of two integral branches: innate and adaptive immunity [1,2]. The innate immune response provides rapid, non-specific defence mechanisms, responding immediately to a wide variety of pathogens [3–5]. In contrast, adaptive immunity is highly specific and builds a memory of prior antigen encounters, allowing for more robust and efficient responses upon re-exposure. Activation of adaptive immunity is dependent on the initial engagement of innate immune responses. Within the adaptive system, B and T-lymphocytes are key players, recognizing antigens, often protein-based, through specialized surface receptors [6]. B-cells, in particular, recognize specific antigenic regions known as epitopes, leading to antibody production. B-cell epitopes are typically found on the surface of antigens and are accessible for direct interaction with B-cell receptors (BCRs) or antibodies, making them essential in shaping the humoral immune response. Identifying B-cell epitopes has broad applications, including understanding pathogen-host interactions, monitoring immune responses, and driving vaccine development [7].

B-cell epitopes are crucial in the design of synthetic vaccines and antibody-based therapies because they represent the specific regions of antigens capable of binding antibodies. Predicting these epitopes allows for a targeted immune response without using the entire antigen, thus enhancing the efficiency of immunotherapeutic. The B-cell epitopes are the clusters of amino acids that are surface accessible. Based on its spatial structure, it can be divided into two groups: linear or sequential (continuous) and non-linear or conformational (discontinuous) [8–11]. Approximately 90% of the B-cell epitopes are discontinuous in nature [9,12]. In the past, several in silico methods have been developed to predict conformational B-cell epitopes. Broadly, these methods can be divided into two categories, sequence-based models and structure-based models. A major limitation of structure-based methods is that they require the structure of the antigen and cannot be used when the structure is not available [13–17]. To overcome this limitation, Ansari and Raghav (2009) developed CBTOPE, a method for predicting conformational B cell epitopes or antibody-interacting residues from the primary sequence of an antigen [18]. In addition, several other methods, such as Epitopia [19], BepiPred 2.0 [20], and SEPIa, [21] are also available for predicting conformational B-cell epitopes in antigens based on their sequence information. In 2023, Cia et al. benchmarked nine state-of-the-art conformational B-cell epitope methods, including both generic and antibody-specific methods, and found that all the methods achieved very low performance, with some performing no better than randomly generated patches of surface residues [22].

A possible explanation of the poor performance of existing methods is that they were trained on small datasets, while the benchmark was performed on large antibody-antigen structure datasets. Thus, there is a need to develop improved versions of existing methods that can predict conformational B-epitopes in a sequence with high accuracy. The "CBTOPE" method has been widely used and extensively cited by the scientific community over the past fifteen years. However, it was originally trained on a small dataset, as only a limited number of antibody-antigen complex structures were available in the Protein Data Bank (PDB) in 2009. In this study, we proposed "CBTOPE2", an improved version of "CBTOPE" that has been trained on a large dataset using state-of-the-art machine learning techniques. We trained, tested, and evaluated our models on a well-curated benchmark dataset from Cia et al. (2023). This updated version outperformed all existing methods tested on the benchmark, including structure-based approaches. We hope that this improved version will be widely adopted by the scientific community, as was our previous version. The system architecture of the method is provided in Figure 1.

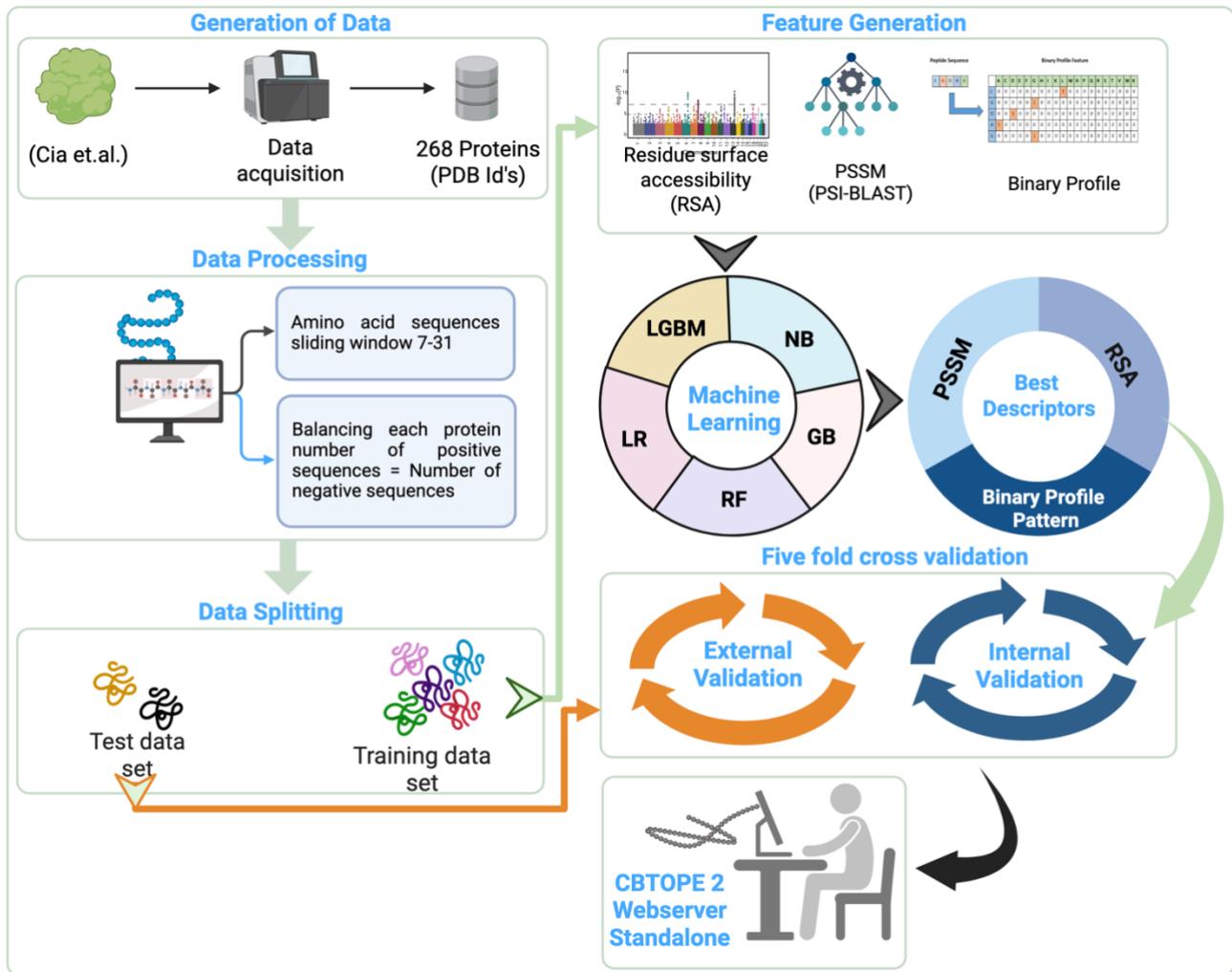

**Figure 1: The complete workflow of the study**

## 3. Materials and Methods
### 3.1. Data Curation
#### 3.1.1. Main Dataset

In this study, we utilized a high-quality dataset curated by Cia et al. (2023) [22]. The dataset comprises 268 non-redundant PDB structures of antibody-antigen complexes, specifically selected for B-cell epitope prediction evaluation. Each structure includes comprehensive residue-level information, distinguishing interacting and non-interacting regions. These structures were obtained from the Protein Data Bank (PDB) and meticulously filtered to ensure high resolution and completeness. We utilized this dataset for both training and testing by splitting it into 80% (214 proteins) for training and 20% (54 proteins) for testing. The dataset files, along with detailed information such as PDB structures and residue labels, are available at https://github.com/3BioCompBio/BCellEpitope .

### 3.1.2. Pattern Formation

The residue interactions are influenced not only by the individual residue itself but also by its nearby residues [23]. Thus, we generated overlapping patterns/sequences of length 7 to 31 residues for all antigens to incorporate this contextual information, as generated in previous studies [18,24–26]. A binary label was assigned to each pattern based on the interaction status of its central residue: positive labels were assigned to interacting residues, negative labels to non-interacting residues. To create the pattern for residues that are present at the terminal ends of sequences, we added a dummy residue.

### 3.1.3. Dataset Balancing

We generated overlapping patterns of each antigen in the training and testing datasets, one pattern for each residue. We assigned a positive pattern if its central residue is antibody-interacting; otherwise, the pattern was assigned as a negative pattern. It was observed that a number of negative patterns is many times more than positive patterns, as only a limited number of residues on an antigen interact with the antibody [18]. We obtained a total of 45,324 negative patterns and 3,980 positive patterns in the training dataset. Similarly, we obtained 11,736 negative patterns and 1,069 positive patterns in the validation dataset. In order to remove the bias of patterns of the negative class, we randomly selected an equal number of negative patterns to positive patterns. Finally, we got 3,980 positive patterns and 3,980 negative patterns in the training dataset. Similarly, we got 1,069 positive and 1,069 negative sequences in the validation dataset.

### 3.3. Feature Generation of Patterns

In order to build an effective antigen-antibody interactions prediction model, the generated patterns were converted into numerical representations using the three types of features: a binary profile, a position-specific scoring matrix (PSSM) [27], and relative solvent accessibility (RSA).

### 3.3.1. Binary Profiles

A binary profile was constructed to represent the amino acid composition of each sequence. This feature was encoded as a 21×W matrix, where 21 corresponds to the 20 standard amino acids plus one dummy residue, and W represents the window size. Each row in this matrix contained binary values (0 or 1), indicating the presence or absence of a particular amino acid at a given position within the sequence [28,29].

### 3.3.2. Position-Specific Scoring Matrix (PSSM)

The PSSM refers to a matrix that holds details about the likelihood of amino acids appearing at specific positions, which is obtained from a multiple sequence alignment. We created the PSSM profiles by utilizing Position-Specific Iterated BLAST (PSI-BLAST) to gather evolutionary data on the protein. This search was conducted against the Swiss Prot Database, applying an E-value threshold of 0.1 and setting the word size to 3, running for three iterations on protein sequences. For the alignment, we employed BLOSUM62 as the scoring matrix, using gap penalties of 11 for opening gaps and 1 for extending gaps [30]. Consequently, we developed PSSM profiles for each antigen in both the training and testing datasets. Following this, we constructed patterns with window lengths ranging from 7 to 31 residues. While forming the feature vector, the dummy variable X is represented as a zero vector of length 20. The final feature vector is organized into a matrix with dimensions Nx20, where N corresponds to the pattern's length [18].

### 3.3.3. Relative Solvent Accessibility (RSA)

The RSA gives a measure of the extent of exposure of a residue in the 3D structure. We have utilized the SPOT-1D, a state-of-the-art tool for predicting protein secondary structure, backbone angles, and solvent accessibility from sequence data [31]. This method uses deep learning models to integrate evolutionary profiles and sequence features, providing highly accurate predictions of residue exposure to the solvent [32]. RSA values were normalized to a scale of 0 to 1, where 0 represents fully buried residues and 1 represents completely exposed residues [33]. SPOT-1D leverages PSSM and other input features to achieve reliable RSA predictions. Only residues with higher RSA values were considered likely candidates for B-cell epitope regions, as surface-exposed residues are more prone to participate in antigen-antibody interactions.

### 3.3.4. PSSM Profile with Binary Profile

The PSSM with binary profile feature set combined evolutionary conservation with sequence identity information. While PSSM captured the likelihood of amino acid substitutions across homologous sequences, the binary profile encoded the presence or absence of specific amino acids at each position within the sequence [18]. This combination allowed the model to simultaneously utilize: (1) PSSM for detecting conserved regions that are functionally critical, and (2) Binary Profile for providing a direct representation of amino acid identity, enabling the model to distinguish between residues with similar conservation scores but different sequence contexts. By integrating these complementary features, the model gained a dual perspective: evolutionary conservation to prioritize important regions and binary sequence identity to refine residue-specific predictions.

### 3.3.5. PSSM with RSA

The PSSM with RSA feature set brought together evolutionary conservation and structural accessibility. RSA values, computed using SPOT-1D [32], provided critical information about the solvent exposure of each residue, highlighting residues more likely to be surface-accessible and available for interactions. This combination enabled the model to utilize: (1) PSSM to detect conserved residues likely important for interactions, and (2) RSA to identify residues with higher solvent accessibility, as surface-exposed residues are more likely to form epitopes or binding interfaces [34]. This feature set was particularly valuable for connecting functional importance PSSM with structural context RSA, enabling the model to better predict interaction-prone residues in antigen-antibody complexes.

### 3.3.6. PSSM with Binary Profile and RSA

The combination of PSSM, binary profile, and RSA feature set represented the most comprehensive combination of features, incorporating evolutionary, sequence-specific, and structural information. This combination was designed to maximize the model's ability to learn patterns relevant to antigen-antibody interactions by leveraging all three dimensions: (1)Evolutionary Information (PSSM) is used to capture the conserved regions that is crucial for interactions [34], and (2) Sequence Identity (Binary Profile) is used to provide precise amino acid composition and sequence patterns, and (3) Structural Exposure (RSA) is used to highlight residues exposed on the protein surface, where interactions are most likely to occur.

By merging these diverse features, this combination provided the richest representation of the antigen sequence, offering insights into both functional and structural determinants of antigen-antibody interactions. This feature set was hypothesized to yield the best performance by leveraging complementary information from multiple sources. The exploration of these feature combinations allowed for a systematic evaluation of how different dimensions of sequence and structure contribute to antigen-antibody interaction prediction. Each combination added a unique perspective to the learning process, with the goal of identifying the most informative feature set for robust epitope prediction.

### 3.4. Model Training

To evaluate the performance of different feature sets in predicting B-cell epitopes, various machine learning models were applied to the curated dataset, each selected for its unique strengths. Random Forest (RF) [35] was included as an ensemble model, combining multiple decision trees to enhance predictive accuracy and mitigate overfitting. Gradient Boosting (GB) was also implemented for its iterative approach in correcting errors made by previous models, contributing to improved accuracy in complex datasets [36]. To balance model complexity, Naive Bayes (NB) was included for its efficiency and speed, particularly useful in handling large datasets, despite assuming feature independence [37]. Additionally, LightGBM (LGBM) was also utilized for its efficiency in handling large datasets with high-dimensional features while maintaining high predictive accuracy through its gradient-based one-side sampling and exclusive feature bundling techniques. Finally, Logistic Regression (LR) was applied due to its interpretability and its strong performance in binary classification tasks, especially in cases where data is linearly separable [38]. Each model's distinct capabilities allowed for a comprehensive evaluation of the dataset, assessing the contribution of various feature sets to predictive performance.

### 3.5. Cross-Validation

In order to train, test, and evaluate our models, we used the standard protocol used in previous studies [39]. All our models were developed using 5-fold cross-validation on the training dataset of 214 antigens. In this technique, one fold is used for testing, and the remaining four folds are used for training. This process is repeated in such a way that each fold is tested once. In order to optimize the parameters of machine learning techniques, performance is optimized on the test dataset. Our final model is evaluated on a validation or independent dataset of 54 antigens, which is not used in the training and testing of models.

### 3.6. Performance Evaluation

In this study, we used both threshold-dependent and threshold-independent parameters to assess the performance of all models [22]. Together, these metrics offered a multidimensional evaluation of model performance, enabling an in-depth comparison across different feature combinations [40]. These parameters were calculated using the following equations:

$$MCC = \frac{(T_P * T_N) - (F_P * F_N)}{\sqrt{(T_P + F_P)(T_P + F_N)(T_N + F_P)(T_N + F_N)}} \quad [1]$$

$$Specificity = \frac{T_N}{T_N + F_P} \quad [2]$$

$$Accuracy = \frac{T_P + T_N}{T_P + T_N + F_P + F_N} \quad [3]$$

$$Sensitivity = \frac{T_P}{T_P + F_N} \quad [4]$$

$$F1 - Score = \frac{2T_P}{2T_P + F_P + F_N} \quad [5]$$

$$Precision = \frac{T_P}{T_P + F_P} \quad [6]$$

Where $T_P$, $T_N$, $F_P$, and $F_N$ stand for true positive, true negative, false positive, and false negative, respectively.

## 4. Results

### 4.1. Machine Learning Models

In this study, we used a wide range of machine learning models to develop a method for predicting conformational B-cell epitopes. These models were developed using features generated from positive and negative patterns. The following are major profiles used to develop the prediction model.

#### 4.1.1. Binary pattern profiles

We developed machine learning models using binary profile patterns (BPP), where each protein/peptide sequence is represented as a 21×W matrix, with W denoting the window size. Models were trained and evaluated across multiple window sizes ranging from 7 to 31. Among the classifiers tested, the LGBM model achieved the best overall performance at a pattern length of 17, with the highest AUROC of 0.58 and MCC of 0.14 on the validation dataset, as shown in Table 1. Other classifiers showed comparable but slightly lower performance. The NB and GB models both achieved an AUROC of 0.57 and an MCC of 0.11 on the validation dataset. The LR model also reached an AUROC of 0.57, but with a lower MCC of 0.10, while the RF model showed the least performance with an AUROC of 0.56 and MCC of 0.08. The complete results across different window sizes are provided in Supplementary File S1.

**Table 1: The ML performance results on window 17 using Binary profile patterns**

| Model | Training Dataset | | | | | Validation Dataset | | | | |
|---|---|---|---|---|---|---|---|---|---|---|
| | Sens | Spec | Acc | AUROC | MCC | Sens | Spec | Acc | AUROC | MCC |
| GB | 0.63 | 0.53 | 0.58 | 0.60 | 0.16 | 0.60 | 0.51 | 0.56 | 0.57 | 0.11 |
| LR | 0.58 | 0.57 | 0.58 | 0.60 | 0.15 | 0.54 | 0.56 | 0.55 | 0.57 | 0.10 |
| NB | 0.58 | 0.57 | 0.57 | 0.60 | 0.15 | 0.56 | 0.55 | 0.56 | 0.57 | 0.11 |
| LGBM | 0.59 | 0.57 | 0.58 | 0.60 | 0.16 | 0.58 | 0.56 | 0.57 | 0.58 | 0.14 |

| RF | 0.60 | 0.54 | 0.57 | 0.59 | 0.13 | 0.56 | 0.52 | 0.54 | 0.56 | 0.08 |

LR, Logistic Regression; RF, Random Forest; NB, Naive Bayes; LGBM, Light Gradient Boosting Machine; GB, Gradient Boost; Sens, Sensitivity; Spec, Specificity; Acc, Accuracy; AUROC, Area Under the Receiver Operating Characteristic; MCC, Matthews Correlation Coefficient

### 4.1.2. Performance on PSSM profiles

In addition to the binary profile, we also used the PSSM (Position-Specific Scoring Matrix) profile for developing prediction models. PSSM profiles, generated using PSI-BLAST, capture evolutionary conservation, making them highly effective for biological sequence classification tasks. A variety of machine learning classifiers were evaluated using PSSM-based features. Among the tested models mentioned in Table 2, RF achieved the highest AUROC of 0.63 on the validation dataset, along with an MCC of 0.16, indicating it as the best-performing model overall. The GB classifier also performed competitively with an AUROC of 0.62 and an MCC of 0.17. NB and LGBM followed closely, both reaching AUROC scores of 0.60, with MCC values of 0.15 and 0.12, respectively. The LR model showed slightly lower performance, achieving an AUROC of 0.59 and an MCC of 0.12. The detailed results across all tested window sizes are provided in Supplementary File S2.

Table 2: The ML performance results on window 17 using the PSSM matrix

| Model | Training Dataset | | | | | Validation Dataset | | | | |
|---|---|---|---|---|---|---|---|---|---|---|
| | Sens | Spec | Acc | AUROC | MCC | Sens | Spec | Acc | AUROC | MCC |
| **RF** | 0.68 | 0.56 | 0.62 | 0.67 | 0.24 | 0.63 | 0.53 | 0.58 | 0.63 | 0.16 |
| **NB** | 0.69 | 0.51 | 0.60 | 0.64 | 0.20 | 0.61 | 0.54 | 0.58 | 0.60 | 0.15 |
| **LR** | 0.63 | 0.57 | 0.60 | 0.65 | 0.20 | 0.59 | 0.52 | 0.56 | 0.59 | 0.12 |
| **GB** | 0.69 | 0.55 | 0.62 | 0.66 | 0.24 | 0.65 | 0.52 | 0.58 | 0.62 | 0.17 |
| **LGBM** | 0.65 | 0.58 | 0.61 | 0.66 | 0.23 | 0.59 | 0.54 | 0.56 | 0.60 | 0.12 |

LR, Logistic Regression; RF, Random Forest; NB, Naive Bayes; LGBM, Light Gradient Boosting Machine; GB, Gradient Boost; Sens, Sensitivity; Spec, Specificity; Acc, Accuracy; AUROC, Area Under the Receiver Operating Characteristic; MCC, Matthews Correlation Coefficient

### 4.1.3. Performance on PSSM with BPP

We also developed models using a combination of PSSM and binary profile features. As shown in Table 3, integrating these two feature types results in improved performance compared to using either profile individually. Among the classifiers evaluated, the RF model achieved the best performance on the validation dataset, with the highest AUROC of 0.63 and MCC of 0.18, indicating strong discriminative capability. The GB classifier reached an AUROC of 0.62 and MCC of 0.17, followed by LGBM with an AUROC of 0.62 and MCC of 0.16. NB and LR obtained AUROC values of 0.60 and 0.59, and MCC scores of 0.15 and 0.11, respectively. These findings suggest that PSSM and binary profile features enhance the predictive power for conformational B-cell epitope identification. The detailed results across different window sizes are provided in Supplementary File S3.

**Table 3: The ML performance results on window 17 using the PSSM matrix with BPP**

| Model | Training Dataset | | | | | Validation Dataset | | | | |
|---|---|---|---|---|---|---|---|---|---|---|
| | Sens | Spec | Acc | AUROC | MCC | Sens | Spec | Acc | AUROC | MCC |
| GB | 0.69 | 0.55 | 0.62 | 0.66 | 0.24 | 0.64 | 0.52 | 0.58 | 0.62 | 0.17 |
| LR | 0.60 | 0.58 | 0.59 | 0.63 | 0.18 | 0.58 | 0.53 | 0.56 | 0.59 | 0.11 |
| NB | 0.65 | 0.54 | 0.60 | 0.63 | 0.19 | 0.60 | 0.55 | 0.58 | 0.60 | 0.15 |
| LGBM | 0.65 | 0.57 | 0.61 | 0.65 | 0.21 | 0.61 | 0.56 | 0.58 | 0.62 | 0.16 |
| RF | 0.67 | 0.55 | 0.61 | 0.66 | 0.22 | 0.64 | 0.54 | 0.59 | 0.63 | 0.18 |

LR, Logistic Regression; RF, Random Forest; NB, Naive Bayes; LGBM, Light Gradient Boosting Machine; GB, Gradient Boost; Sens, Sensitivity; Spec, Specificity; Acc, Accuracy; AUROC, Area Under the Receiver Operating Characteristic; MCC, Matthews Correlation Coefficient;

### 4.1.4. Performance on PSSM with RSA

We also developed models using a combination of PSSM and Relative Surface Accessibility (RSA) profiles to predict conformational B-cell epitopes. As shown in Table 4, incorporating RSA alongside evolutionary information from PSSM improved predictive performance. The RF model performed well among the classifiers, achieving the highest AUROC of 0.64 and an MCC of 0.18 on the validation dataset, highlighting its strong capability in distinguishing interacting from non-interacting residues. Other classifiers also demonstrated competitive performance. The LR attained an AUROC of 0.60 and an MCC of 0.14, while LGBM achieved an AUROC of 0.62 and an MCC of 0.17. The NB model showed moderate performance with an AUROC of 0.61 and an MCC of 0.18, whereas GB achieved an AUROC of 0.62 and an MCC of 0.19. These results support the notion that RSA adds valuable structural context to evolutionary profiles, enhancing model performance for epitope prediction. Detailed performance metrics across additional window sizes are provided in Supplementary File S4.

**Table 4: The ML performance results on window 17 using PSSM matrix with RSA features**

| Model | Training Dataset | | | | | Validation Dataset | | | | |
|---|---|---|---|---|---|---|---|---|---|---|
| | Sens | Spec | Acc | AUROC | MCC | Sens | Spec | Acc | AUROC | MCC |
| GB | 0.70 | 0.55 | 0.63 | 0.67 | 0.25 | 0.65 | 0.53 | 0.59 | 0.62 | 0.19 |
| LR | 0.64 | 0.59 | 0.62 | 0.66 | 0.23 | 0.58 | 0.55 | 0.57 | 0.60 | 0.14 |
| NB | 0.73 | 0.47 | 0.60 | 0.64 | 0.21 | 0.65 | 0.53 | 0.59 | 0.61 | 0.18 |
| LGBM | 0.65 | 0.58 | 0.62 | 0.66 | 0.23 | 0.63 | 0.55 | 0.59 | 0.62 | 0.17 |
| RF | 0.67 | 0.56 | 0.61 | 0.66 | 0.23 | 0.64 | 0.54 | 0.59 | 0.64 | 0.18 |

### 4.1.5. Performance on PSSM with BPP and RSA

Similarly, we also combined PSSM and RSA with the Binary Profile to develop prediction models [https://doi.org/10.1155/2014/689219]. As shown in Table 5, incorporating all three features, PSSM, RSA, and Binary Profile, resulted in performance comparable to using only PSSM and RSA. The RF model remained the top performer, achieving an AUROC of 0.63 and an MCC of 0.18 on the validation dataset, reflecting its ability to effectively distinguish interacting and non-interacting. The GB and LGBM also demonstrated strong performance, with an AUROC of 0.62 each and an MCC of 0.18 and 0.17, respectively. Whereas LR and NB produced modest results, with AUROC scores of 0.60 each and MCC values of 0.12 and 0.16, respectively. These results suggest that while the inclusion of Binary Profile alongside RSA and PSSM contributes marginally to overall performance, the structural (RSA) and evolutionary (PSSM) features remain the most influential for epitope prediction. Comprehensive results across all tested window sizes are provided in Supplementary File S5.

**Table 5: The ML performance results on window 17 using PSSM matrix with RSA and Binary profile**

| Model | Training Dataset | | | | | Validation Dataset | | | | |
|---|---|---|---|---|---|---|---|---|---|---|
| | Sens | Spec | Acc | AUROC | MCC | Sens | Spec | Acc | AUROC | MCC |
| **GB** | 0.70 | 0.55 | 0.62 | 0.67 | 0.25 | 0.65 | 0.52 | 0.59 | **0.62** | 0.18 |
| **LR** | 0.61 | 0.59 | 0.60 | 0.65 | 0.20 | 0.58 | 0.54 | 0.56 | **0.60** | 0.12 |
| **NB** | 0.68 | 0.52 | 0.60 | 0.64 | 0.21 | 0.62 | 0.54 | 0.58 | **0.60** | 0.16 |
| **LGBM** | 0.65 | 0.58 | 0.61 | 0.66 | 0.23 | 0.61 | 0.56 | 0.58 | **0.62** | 0.17 |
| **RF** | 0.67 | 0.57 | 0.62 | 0.67 | 0.24 | 0.63 | 0.55 | 0.59 | **0.63** | 0.18 |

LR, Logistic Regression; RF, Random Forest; NB, Naive Bayes; LGBM, Light Gradient Boosting Machine; GB, Gradient Boost; Sens, Sensitivity; Spec, Specificity; Acc, Accuracy; AUROC, Area Under the Receiver Operating Characteristic; MCC, Matthews Correlation Coefficient;

## 4.3. Final Model

Based on the evaluation of different feature combinations and machine learning models, RF trained on PSSM and RSA emerged as the best-performing model. It achieved the highest AUROC of 0.64 and MCC of 0.18, demonstrating its effectiveness in capturing key structural and evolutionary patterns in antigen-antibody interactions. While the Binary Profile feature had minimal impact, RSA significantly enhanced model performance by providing insights into residue solvent exposure. The results confirm that structural accessibility (RSA) and evolutionary conservation

(PSSM) are the most informative features for B-cell epitope prediction. Thus, the final model prioritizes PSSM with RSA with RF for optimal predictive performance.

## 5. Web Server

To aid the scientific community, we have created a web server (https://webs.iiitd.edu.in/raghava/cbtope2) that is easy to use and accessible, featuring the best prediction models from this research [20]. Researchers can predict which residues of antibodies will bind to antigen sequences by simply uploading their antigen data in FASTA format. The server processes the entered antigens using the optimized model to find possible antibody-binding locations. The findings are organized clearly, showing the antibody-binding residues in the structure of the antigen. Furthermore, users can download the datasets that were used for both training and testing in this research. The web server, built with HTML, JavaScript, and PHP, works on various devices, including laptops, Android phones, iPhones, and iPads. To improve accessibility, we also provided a standalone version of CBTOPE2, allowing researchers to carry out predictions offline without needing an internet connection. This standalone software is available on GitHub (https://github.com/raghavagps/cbtope2) and can be installed as a PyPI package using pip (pip install cbtope2). Both the web server and the standalone software are designed to make advanced computational tools more available, promoting advancements in the understanding of antigen-antibody interactions. The web server can be found at https://webs.iiitd.edu.in/raghava/cbtope2/, serving as a dependable source for the scientific community.

## 6. Discussion and Conclusion

This study underscores the significance of integrating sequence-based and structural features for improved B-cell epitope prediction. The incorporation of PSSM and RSA into the predictive framework led to notable performance enhancements, with Random Forest achieving an MCC of 0.18 and an AUROC of 0.64. RSA, which captures solvent accessibility, emerged as a key determinant in identifying surface-exposed residues likely to interact with antibodies. This finding highlights the critical role of structural information in complementing sequence-based features for predictive modeling. Although binary profiles did not contribute, but RSA contributed minimally, its combination with PSSM yielded slight improvements in performance, emphasizing the value of integrating diverse feature sets. Among the models tested, Random Forest consistently demonstrated superior performance by effectively capturing complex feature interactions, while simpler models like logistic regression occasionally achieved comparable results under certain conditions. The use of balanced datasets in this study ensured an unbiased evaluation of the models, enhancing the reliability of the findings. These results emphasize the importance of feature-rich models for achieving higher predictive accuracy and biological relevance. Future research should focus on leveraging larger and more diverse datasets, advanced feature engineering techniques, and hybrid modeling approaches to further refine B-cell epitope prediction. Such advancements hold promise for enabling breakthroughs in immunotherapeutics, vaccine development, and antibody engineering.

**Funding Source**

The current work has been supported by the Department of Biotechnology (DBT) grant BT/PR40158/BTIS/137/24/2021.

**Conflict of interest**

The authors declare no competing financial and non-financial interests.

**Authors' contributions**

GPSR collected the dataset. AP, M, and NK processed the dataset. AP, M, and GPSR implemented the algorithms and developed the prediction models. NK, M, AP, and GPSR analysed the results. M and AP created the front-end, back-end of the webserver and standalone. RS contributed to resolution of server-side integration issues, and domain-specific clarification of biological concepts. AP, M, RS, NK, and GPSR penned the manuscript. GPSR conceived and coordinated the project. All authors have read and approved the final manuscript.

# Acknowledgments

Authors are thankful to the University Grants Commission (UGC) and DST-Inspire (KVPY), and IIIT-Delhi for fellowships and financial support, and the Department of Computational Biology, IIITD New Delhi for infrastructure and facilities. We would like to acknowledge that Figures were created using BioRender.com.

# Data Availability Statement

All the datasets used in this study are available in MAP (Modification and Annotation in Proteins) [41] format at the "CBTOPE2" web server, at https://webs.iiitd.edu.in/raghava/cbtope2/dataset.html.

# References

[1] J. Parkin, B. Cohen, An overview of the immune system, Lancet 357 (2001) 1777–1789. https://doi.org/10.1016/S0140-6736(00)04904-7.

[2] H.R. Ansari, D.R. Flower, G.P.S. Raghava, AntigenDB: an immunoinformatics database of pathogen antigens, Nucleic Acids Res. 38 (2010) D847-53. https://doi.org/10.1093/nar/gkp830.

[3] G. Nagpal, K. Chaudhary, P. Agrawal, G.P.S. Raghava, Computer-aided prediction of antigen presenting cell modulators for designing peptide-based vaccine adjuvants, J. Transl. Med. 16 (2018) 181. https://doi.org/10.1186/s12967-018-1560-1.

[4] D. Kaur, C. Arora, G.P.S. Raghava, A hybrid model for predicting pattern recognition receptors using evolutionary information, Front. Immunol. 11 (2020) 71. https://doi.org/10.3389/fimmu.2020.00071.

[5] D. Kaur, S. Patiyal, N. Sharma, S.S. Usmani, G.P.S. Raghava, PRRDB 2.0: a comprehensive database of pattern-recognition receptors and their ligands, Database (Oxford) 2019 (2019). https://doi.org/10.1093/database/baz076.

[6] S. Lata, M. Bhasin, G.P.S. Raghava, Application of machine learning techniques in predicting MHC binders, Methods Mol. Biol. 409 (2007) 201–215. https://doi.org/10.1007/978-1-60327-118-9_14.

[7] S.E.C. Caoili, Comprehending B-cell Epitope prediction to develop vaccines and immunodiagnostics, Front. Immunol. 13 (2022) 908459. https://doi.org/10.3389/fimmu.2022.908459.

[8] M.Z. Atassi, J.A. Smith, A proposal for the nomenclature of antigenic sites in peptides and proteins, Immunochemistry 15 (1978) 609–610. https://doi.org/10.1016/0161-5890(78)90016-0.

[9] M.C. Jespersen, S. Mahajan, B. Peters, M. Nielsen, P. Marcatili, Antibody specific B-cell Epitope predictions: Leveraging information from antibody-antigen protein complexes, Front. Immunol. 10 (2019) 298. https://doi.org/10.3389/fimmu.2019.00298.

[10] L. Potocnakova, M. Bhide, L.B. Pulzova, An introduction to B-cell Epitope mapping and in silico Epitope prediction, J. Immunol. Res. 2016 (2016) 6760830. https://doi.org/10.1155/2016/6760830.

[11] A. Bhardwaj, V. Scaria, G.P.S. Raghava, A.M. Lynn, N. Chandra, S. Banerjee, M.V. Raghunandanan, V. Pandey, B. Taneja, J. Yadav, D. Dash, J. Bhattacharya, A. Misra, A. Kumar, S. Ramachandran, Z. Thomas, Open Source Drug Discovery Consortium, S.K. Brahmachari, Open source drug discovery--a new paradigm of collaborative research in tuberculosis drug development, Tuberculosis (Edinb.) 91 (2011) 479–486. https://doi.org/10.1016/j.tube.2011.06.004.

[12] S. Ferdous, S. Kelm, T.S. Baker, J. Shi, A.C.R. Martin, B-cell epitopes: Discontinuity and conformational analysis, Mol. Immunol. 114 (2019) 643–650. https://doi.org/10.1016/j.molimm.2019.09.014.

[13] U. Kulkarni-Kale, S. Bhosle, A.S. Kolaskar, CEP: a conformational epitope prediction server, Nucleic Acids Res. 33 (2005) W168-71. https://doi.org/10.1093/nar/gki460.

[14] J.V. Kringelum, C. Lundegaard, O. Lund, M. Nielsen, Reliable B cell epitope predictions: impacts of method development and improved benchmarking, PLoS Comput. Biol. 8 (2012) e1002829. https://doi.org/10.1371/journal.pcbi.1002829.

[15] T. Qi, T. Qiu, Q. Zhang, K. Tang, Y. Fan, J. Qiu, D. Wu, W. Zhang, Y. Chen, J. Gao, R. Zhu, Z. Cao, SEPPA 2.0--more refined server to predict spatial epitope considering species of immune host and subcellular localization of protein antigen, Nucleic Acids Res. 42 (2014) W59-63. https://doi.org/10.1093/nar/gku395.

[16] B.M. da Silva, Y. Myung, D.B. Ascher, D.E.V. Pires, epitope3D: a machine learning method for conformational B-cell epitope prediction, Brief. Bioinform. 23 (2022). https://doi.org/10.1093/bib/bbab423.

[17] B. Solihah, A. Azhari, A. Musdholifah, Enhancement of conformational B-cell epitope prediction using CluSMOTE, PeerJ Comput. Sci. 6 (2020) e275. https://doi.org/10.7717/peerj-cs.275.
[18] H.R. Ansari, G.P. Raghava, Identification of conformational B-cell Epitopes in an antigen from its primary sequence, Immunome Res. 6 (2010) 6. https://doi.org/10.1186/1745-7580-6-6.
[19] N.D. Rubinstein, I. Mayrose, E. Martz, T. Pupko, Epitopia: a web-server for predicting B-cell epitopes, BMC Bioinformatics 10 (2009) 287. https://doi.org/10.1186/1471-2105-10-287.
[20] M.C. Jespersen, B. Peters, M. Nielsen, P. Marcatili, BepiPred-2.0: improving sequence-based B-cell epitope prediction using conformational epitopes, Nucleic Acids Res. 45 (2017) W24–W29. https://doi.org/10.1093/nar/gkx346.
[21] G.A. Dalkas, M. Rooman, SEPIa, a knowledge-driven algorithm for predicting conformational B-cell epitopes from the amino acid sequence, BMC Bioinformatics 18 (2017) 95. https://doi.org/10.1186/s12859-017-1528-9.
[22] G. Cia, F. Pucci, M. Rooman, Critical review of conformational B-cell epitope prediction methods, Brief. Bioinform. 24 (2023). https://doi.org/10.1093/bib/bbac567.
[23] J. Garnier, J.F. Gibrat, B. Robson, GOR method for predicting protein secondary structure from amino acid sequence, Methods Enzymol. 266 (1996) 540–553. https://doi.org/10.1016/s0076-6879(96)66034-0.
[24] H. Kaur, G.P.S. Raghava, A neural-network based method for prediction of gamma-turns in proteins from multiple sequence alignment, Protein Sci. 12 (2003) 923–929. https://doi.org/10.1110/ps.0241703.
[25] H. Kaur, G.P.S. Raghava, Prediction of alpha-turns in proteins using PSI-BLAST profiles and secondary structure information, Proteins 55 (2004) 83–90. https://doi.org/10.1002/prot.10569.
[26] J.S. Chauhan, A.H. Bhat, G.P.S. Raghava, A. Rao, GlycoPP: a webserver for prediction of N- and O-glycosites in prokaryotic protein sequences, PLoS One 7 (2012) e40155. https://doi.org/10.1371/journal.pone.0040155.
[27] S.Y.-H. Lin, C.-W. Cheng, E.C.-Y. Su, Prediction of B-cell epitopes using evolutionary information and propensity scales, BMC Bioinformatics 14 Suppl 2 (2013) S10. https://doi.org/10.1186/1471-2105-14-s2-s10.
[28] X. Xiao, P. Wang, K.-C. Chou, GPCR-CA: A cellular automaton image approach for predicting G-protein-coupled receptor functional classes, J. Comput. Chem. 30 (2009) 1414–1423. https://doi.org/10.1002/jcc.21163.
[29] X. Xiao, S. Shao, Y. Ding, Z. Huang, K.-C. Chou, Using cellular automata images and pseudo amino acid composition to predict protein subcellular location, Amino Acids 30 (2006) 49–54. https://doi.org/10.1007/s00726-005-0225-6.
[30] S.F. Altschul, T.L. Madden, A.A. Schäffer, J. Zhang, Z. Zhang, W. Miller, D.J. Lipman, Gapped BLAST and PSI-BLAST: a new generation of protein database search programs, Nucleic Acids Res. 25 (1997) 3389–3402. https://doi.org/10.1093/nar/25.17.3389.
[31] R. Heffernan, Y. Yang, K. Paliwal, Y. Zhou, Capturing non-local interactions by long short-term memory bidirectional recurrent neural networks for improving prediction of protein secondary structure, backbone angles, contact numbers and solvent accessibility, Bioinformatics 33 (2017) 2842–2849. https://doi.org/10.1093/bioinformatics/btx218.
[32] J. Singh, T. Litfin, K. Paliwal, J. Singh, A.K. Hanumanthappa, Y. Zhou, SPOT-1D-Single: improving the single-sequence-based prediction of protein secondary structure, backbone angles, solvent accessibility and half-sphere exposures using a large training set and ensembled deep learning, Bioinformatics 37 (2021) 3464–3472. https://doi.org/10.1093/bioinformatics/btab316.
[33] M.Z. Tien, A.G. Meyer, D.K. Sydykova, S.J. Spielman, C.O. Wilke, Maximum allowed solvent accessibilites of residues in proteins, PLoS One 8 (2013) e80635. https://doi.org/10.1371/journal.pone.0080635.

[34] J. Zhang, X. Zhao, P. Sun, B. Gao, Z. Ma, Conformational B-cell epitopes prediction from sequences using cost-sensitive ensemble classifiers and spatial clustering, Biomed Res. Int. 2014 (2014) 689219. https://doi.org/10.1155/2014/689219.

[35] L. Breiman, Mach. Learn. 45 (2001) 5–32. https://doi.org/10.1023/a:1010933404324.

[36] J.H. Friedman, Greedy function approximation: A gradient boosting machine, Ann. Stat. 29 (2001) 1189–1232. https://doi.org/10.1214/aos/1013203451.

[37] G.I. Webb, Naïve Bayes, in: Encyclopedia of Machine Learning, Springer US, Boston, MA, 2011: pp. 713–714. https://doi.org/10.1007/978-0-387-30164-8_576.

[38] C.-Y.J. Peng, K.L. Lee, G.M. Ingersoll, An introduction to logistic regression analysis and reporting, J. Educ. Res. 96 (2002) 3–14. https://doi.org/10.1080/00220670209598786.

[39] S. Saha, G.P.S. Raghava, Prediction of continuous B-cell epitopes in an antigen using recurrent neural network, Proteins 65 (2006) 40–48. https://doi.org/10.1002/prot.21078.

[40] M. Sokolova, G. Lapalme, A systematic analysis of performance measures for classification tasks, Inf. Process. Manag. 45 (2009) 427–437. https://doi.org/10.1016/j.ipm.2009.03.002.

[41] A. Shendre, N.K. Mehta, A.S. Rathore, N. Kumar, S. Patiyal, G.P.S. Raghava, MAP format for representing chemical modifications, annotations, and mutations in protein sequences: An extension of the FASTA format, ArXiv [q-Bio.BM] (2025). https://doi.org/10.48550/ARXIV.2505.03403.

# SUPPLEMENTARY TABLES

## Supplementary Table S1: The complete result on different windows using binary profile

LR, Logistic Regression; RF, Random Forest; NB, Naive Bayes; LGBM, Light Gradient Boosting Machine; GB, Gradient Boost; Sens, Sensitivity; Spec, Specificity; Acc, Accuracy; AUROC, Area Under the Receiver Operating Characteristic; MCC, Matthews Correlation Coefficient

### Window 7

| Model | Training Dataset | | | | | Validation dataset | | | | |
|---|---|---|---|---|---|---|---|---|---|---|
| | Sens | Spec | Acc | AUROC | MCC | Sens | Spec | Acc | AUROC | MCC |
| RF | 0.59 | 0.54 | 0.56 | 0.59 | 0.13 | 0.58 | 0.54 | 0.56 | 0.58 | 0.12 |
| LGBM | 0.58 | 0.57 | 0.57 | 0.61 | 0.15 | 0.55 | 0.56 | 0.55 | 0.58 | 0.11 |
| NB | 0.59 | 0.57 | 0.58 | 0.61 | 0.16 | 0.58 | 0.55 | 0.56 | 0.59 | 0.13 |
| LR | 0.59 | 0.57 | 0.58 | 0.61 | 0.16 | 0.56 | 0.55 | 0.55 | 0.59 | 0.11 |
| GB | 0.62 | 0.54 | 0.58 | 0.61 | 0.16 | 0.61 | 0.52 | 0.57 | 0.59 | 0.13 |

### Window 9

| Model | Training Dataset | | | | | Validation dataset | | | | |
|---|---|---|---|---|---|---|---|---|---|---|
| | Sens | Spec | Acc | AUROC | MCC | Sens | Spec | Acc | AUROC | MCC |
| RF | 0.59 | 0.54 | 0.56 | 0.59 | 0.13 | 0.57 | 0.54 | 0.56 | 0.58 | 0.12 |
| LGBM | 0.59 | 0.56 | 0.58 | 0.61 | 0.15 | 0.56 | 0.58 | 0.57 | 0.59 | 0.14 |
| NB | 0.59 | 0.57 | 0.58 | 0.61 | 0.16 | 0.56 | 0.57 | 0.57 | 0.59 | 0.13 |
| LR | 0.59 | 0.57 | 0.58 | 0.61 | 0.16 | 0.55 | 0.58 | 0.56 | 0.59 | 0.12 |
| GB | 0.63 | 0.53 | 0.58 | 0.61 | 0.17 | 0.62 | 0.52 | 0.57 | 0.59 | 0.14 |

### Window 11

| Model | Training Dataset | | | | | Validation dataset | | | | |
|---|---|---|---|---|---|---|---|---|---|---|
| | Sens | Spec | Acc | AUROC | MCC | Sens | Spec | Acc | AUROC | MCC |
| RF | 0.60 | 0.53 | 0.57 | 0.59 | 0.13 | 0.59 | 0.54 | 0.57 | 0.59 | 0.13 |
| LGBM | 0.59 | 0.57 | 0.58 | 0.60 | 0.15 | 0.56 | 0.57 | 0.57 | 0.59 | 0.13 |
| NB | 0.59 | 0.57 | 0.58 | 0.60 | 0.15 | 0.58 | 0.57 | 0.57 | 0.59 | 0.14 |
| LR | 0.59 | 0.57 | 0.58 | 0.60 | 0.16 | 0.55 | 0.58 | 0.57 | 0.59 | 0.13 |
| GB | 0.63 | 0.53 | 0.58 | 0.60 | 0.16 | 0.62 | 0.52 | 0.57 | 0.59 | 0.14 |

### Window 13

| Model | Training Dataset | | | | | Validation dataset | | | | |
|---|---|---|---|---|---|---|---|---|---|---|
| | Sens | Spec | Acc | AUROC | MCC | Sens | Spec | Acc | AUROC | MCC |
| RF | 0.59 | 0.53 | 0.56 | 0.58 | 0.13 | 0.60 | 0.55 | 0.57 | 0.59 | 0.15 |
| LGBM | 0.58 | 0.55 | 0.56 | 0.59 | 0.13 | 0.58 | 0.55 | 0.56 | 0.59 | 0.13 |
| NB | 0.58 | 0.57 | 0.57 | 0.60 | 0.15 | 0.56 | 0.55 | 0.56 | 0.58 | 0.12 |
| LR | 0.58 | 0.58 | 0.58 | 0.60 | 0.16 | 0.55 | 0.56 | 0.55 | 0.58 | 0.11 |
| GB | 0.61 | 0.53 | 0.57 | 0.59 | 0.14 | 0.60 | 0.51 | 0.56 | 0.58 | 0.11 |

## Window 15

| Model | Training Dataset | | | | | Validation dataset | | | | |
|---|---|---|---|---|---|---|---|---|---|---|
| | Sens | Spec | Acc | AUROC | MCC | Sens | Spec | Acc | AUROC | MCC |
| RF | 0.59 | 0.54 | 0.56 | 0.59 | 0.13 | 0.58 | 0.52 | 0.55 | 0.58 | 0.10 |
| LGBM | 0.59 | 0.57 | 0.58 | 0.60 | 0.17 | 0.57 | 0.54 | 0.55 | 0.58 | 0.11 |
| NB | 0.57 | 0.57 | 0.57 | 0.60 | 0.15 | 0.58 | 0.54 | 0.56 | 0.58 | 0.12 |
| LR | 0.58 | 0.57 | 0.58 | 0.60 | 0.15 | 0.54 | 0.55 | 0.54 | 0.57 | 0.09 |
| GB | 0.61 | 0.54 | 0.57 | 0.60 | 0.15 | 0.61 | 0.51 | 0.56 | 0.58 | 0.12 |

## Window 17

| Model | Training Dataset | | | | | Validation dataset | | | | |
|---|---|---|---|---|---|---|---|---|---|---|
| | Sens | Spec | Acc | AUROC | MCC | Sens | Spec | Acc | AUROC | MCC |
| RF | 0.60 | 0.54 | 0.57 | 0.59 | 0.13 | 0.56 | 0.52 | 0.54 | 0.56 | 0.08 |
| LGBM | 0.59 | 0.57 | 0.58 | 0.60 | 0.16 | 0.58 | 0.56 | 0.57 | 0.58 | 0.14 |
| NB | 0.58 | 0.57 | 0.57 | 0.60 | 0.15 | 0.56 | 0.55 | 0.56 | 0.57 | 0.11 |
| LR | 0.58 | 0.57 | 0.58 | 0.60 | 0.15 | 0.54 | 0.56 | 0.55 | 0.57 | 0.10 |
| GB | 0.63 | 0.53 | 0.58 | 0.60 | 0.16 | 0.60 | 0.51 | 0.56 | 0.57 | 0.11 |

## Window 19

| Model | Training Dataset | | | | | Validation dataset | | | | |
|---|---|---|---|---|---|---|---|---|---|---|
| | Sens | Spec | Acc | AUROC | MCC | Sens | Spec | Acc | AUROC | MCC |
| RF | 0.58 | 0.52 | 0.55 | 0.58 | 0.10 | 0.57 | 0.54 | 0.56 | 0.58 | 0.11 |
| LGBM | 0.58 | 0.57 | 0.57 | 0.60 | 0.15 | 0.55 | 0.55 | 0.55 | 0.58 | 0.11 |
| NB | 0.57 | 0.57 | 0.57 | 0.59 | 0.13 | 0.56 | 0.56 | 0.56 | 0.58 | 0.11 |
| LR | 0.57 | 0.58 | 0.57 | 0.59 | 0.14 | 0.54 | 0.56 | 0.55 | 0.57 | 0.11 |
| GB | 0.61 | 0.53 | 0.57 | 0.59 | 0.14 | 0.60 | 0.51 | 0.55 | 0.58 | 0.11 |

## Window 21

| Model | Training Dataset | | | | | Validation dataset | | | | |
|---|---|---|---|---|---|---|---|---|---|---|
| | Sens | Spec | Acc | AUROC | MCC | Sens | Spec | Acc | AUROC | MCC |
| RF | 0.60 | 0.52 | 0.56 | 0.58 | 0.12 | 0.55 | 0.54 | 0.55 | 0.57 | 0.10 |
| LGBM | 0.60 | 0.56 | 0.58 | 0.60 | 0.16 | 0.56 | 0.55 | 0.56 | 0.58 | 0.11 |
| NB | 0.57 | 0.57 | 0.57 | 0.59 | 0.14 | 0.54 | 0.54 | 0.54 | 0.57 | 0.08 |
| LR | 0.57 | 0.57 | 0.57 | 0.60 | 0.14 | 0.53 | 0.55 | 0.54 | 0.56 | 0.08 |
| GB | 0.61 | 0.53 | 0.57 | 0.59 | 0.14 | 0.61 | 0.51 | 0.56 | 0.57 | 0.12 |

## Window 23

| Model | Training Dataset | | | | | Validation dataset | | | | |
|---|---|---|---|---|---|---|---|---|---|---|
| | Sens | Spec | Acc | AUROC | MCC | Sens | Spec | Acc | AUROC | MCC |
| RF | 0.59 | 0.53 | 0.56 | 0.58 | 0.12 | 0.56 | 0.53 | 0.54 | 0.57 | 0.09 |
| LGBM | 0.59 | 0.56 | 0.58 | 0.60 | 0.15 | 0.56 | 0.54 | 0.55 | 0.58 | 0.10 |

| | Sens | Spec | Acc | AUROC | MCC | Sens | Spec | Acc | AUROC | MCC |
|---|---|---|---|---|---|---|---|---|---|---|
| **NB** | 0.58 | 0.56 | 0.57 | 0.59 | 0.14 | 0.56 | 0.54 | 0.55 | 0.57 | 0.10 |
| **LR** | 0.57 | 0.57 | 0.57 | 0.59 | 0.14 | 0.53 | 0.54 | 0.54 | 0.57 | 0.08 |
| **GB** | 0.62 | 0.53 | 0.57 | 0.59 | 0.15 | 0.62 | 0.50 | 0.56 | 0.58 | 0.12 |
| | | | | | | | | | | |
| | colspan | | | | | | | | | |

| | Window 25 | | | | | | | | | |
|---|---|---|---|---|---|---|---|---|---|---|
| **Model** | **Training Dataset** | | | | | **Validation dataset** | | | | |
| | Sens | Spec | Acc | AUROC | MCC | Sens | Spec | Acc | AUROC | MCC |
| **RF** | 0.59 | 0.52 | 0.56 | 0.58 | 0.11 | 0.58 | 0.51 | 0.55 | 0.57 | 0.09 |
| **LGBM** | 0.59 | 0.55 | 0.57 | 0.59 | 0.14 | 0.57 | 0.52 | 0.54 | 0.56 | 0.09 |
| **NB** | 0.57 | 0.55 | 0.56 | 0.58 | 0.12 | 0.57 | 0.52 | 0.55 | 0.56 | 0.10 |
| **LR** | 0.56 | 0.56 | 0.56 | 0.59 | 0.13 | 0.55 | 0.54 | 0.54 | 0.56 | 0.09 |
| **GB** | 0.60 | 0.52 | 0.56 | 0.58 | 0.12 | 0.60 | 0.51 | 0.55 | 0.57 | 0.11 |

| | Window 27 | | | | | | | | | |
|---|---|---|---|---|---|---|---|---|---|---|
| **Model** | **Training Dataset** | | | | | **Validation dataset** | | | | |
| | Sens | Spec | Acc | AUROC | MCC | Sens | Spec | Acc | AUROC | MCC |
| **RF** | 0.57 | 0.52 | 0.54 | 0.57 | 0.09 | 0.57 | 0.52 | 0.55 | 0.56 | 0.09 |
| **LGBM** | 0.59 | 0.55 | 0.57 | 0.60 | 0.14 | 0.58 | 0.54 | 0.56 | 0.58 | 0.12 |
| **NB** | 0.56 | 0.56 | 0.56 | 0.58 | 0.12 | 0.56 | 0.53 | 0.54 | 0.57 | 0.09 |
| **LR** | 0.56 | 0.56 | 0.56 | 0.58 | 0.12 | 0.53 | 0.55 | 0.54 | 0.56 | 0.08 |
| **GB** | 0.61 | 0.52 | 0.57 | 0.59 | 0.13 | 0.62 | 0.49 | 0.55 | 0.57 | 0.11 |

| | Window 29 | | | | | | | | | |
|---|---|---|---|---|---|---|---|---|---|---|
| **Model** | **Training Dataset** | | | | | **Validation dataset** | | | | |
| | Sens | Spec | Acc | AUROC | MCC | Sens | Spec | Acc | AUROC | MCC |
| **RF** | 0.59 | 0.52 | 0.56 | 0.57 | 0.11 | 0.60 | 0.50 | 0.55 | 0.57 | 0.11 |
| **LGBM** | 0.58 | 0.55 | 0.57 | 0.59 | 0.13 | 0.60 | 0.53 | 0.56 | 0.58 | 0.13 |
| **NB** | 0.57 | 0.54 | 0.55 | 0.58 | 0.11 | 0.59 | 0.53 | 0.56 | 0.58 | 0.11 |
| **LR** | 0.56 | 0.56 | 0.56 | 0.58 | 0.12 | 0.56 | 0.55 | 0.55 | 0.57 | 0.10 |
| **GB** | 0.61 | 0.52 | 0.57 | 0.59 | 0.14 | 0.62 | 0.49 | 0.55 | 0.57 | 0.11 |

| | Window 31 | | | | | | | | | |
|---|---|---|---|---|---|---|---|---|---|---|
| **Model** | **Training Dataset** | | | | | **Validation dataset** | | | | |
| | Sens | Spec | Acc | AUROC | MCC | Sens | Spec | Acc | AUROC | MCC |
| **RF** | 0.58 | 0.53 | 0.55 | 0.58 | 0.11 | 0.58 | 0.52 | 0.55 | 0.57 | 0.10 |
| **LGBM** | 0.58 | 0.56 | 0.57 | 0.59 | 0.14 | 0.58 | 0.52 | 0.55 | 0.58 | 0.10 |
| **NB** | 0.58 | 0.56 | 0.57 | 0.59 | 0.14 | 0.56 | 0.55 | 0.56 | 0.58 | 0.12 |
| **LR** | 0.56 | 0.56 | 0.56 | 0.58 | 0.12 | 0.56 | 0.54 | 0.55 | 0.57 | 0.10 |
| **GB** | 0.59 | 0.52 | 0.56 | 0.58 | 0.12 | 0.61 | 0.50 | 0.56 | 0.58 | 0.11 |

## Supplementary Table S2: The complete result on different windows using PSSM profile

LR, Logistic Regression; RF, Random Forest; NB, Naive Bayes; LGBM, Light Gradient Boosting Machine; GB, Gradient Boost; Sens, Sensitivity; Spec, Specificity; Acc, Accuracy; AUROC, Area Under the Receiver Operating Characteristic; MCC, Matthews Correlation Coefficient

### Window 7

| Model | Training Dataset | | | | | Validation dataset | | | | |
|---|---|---|---|---|---|---|---|---|---|---|
| | Sens | Spec | Acc | AUROC | MCC | Sens | Spec | Acc | AUROC | MCC |
| NB | 0.71 | 0.51 | 0.61 | 0.64 | 0.22 | 0.67 | 0.53 | 0.60 | 0.62 | 0.20 |
| RF | 0.68 | 0.56 | 0.62 | 0.67 | 0.24 | 0.65 | 0.55 | 0.60 | 0.64 | 0.20 |
| LR | 0.65 | 0.56 | 0.61 | 0.65 | 0.21 | 0.64 | 0.55 | 0.59 | 0.63 | 0.18 |
| GB | 0.69 | 0.54 | 0.62 | 0.66 | 0.23 | 0.69 | 0.52 | 0.61 | 0.64 | 0.21 |
| LGBM | 0.66 | 0.57 | 0.61 | 0.66 | 0.23 | 0.65 | 0.55 | 0.60 | 0.63 | 0.20 |

### Window 9

| Model | Training Dataset | | | | | Validation dataset | | | | |
|---|---|---|---|---|---|---|---|---|---|---|
| | Sens | Spec | Acc | AUROC | MCC | Sens | Spec | Acc | AUROC | MCC |
| NB | 0.70 | 0.50 | 0.60 | 0.64 | 0.21 | 0.64 | 0.53 | 0.59 | 0.62 | 0.18 |
| RF | 0.67 | 0.56 | 0.61 | 0.67 | 0.23 | 0.65 | 0.54 | 0.59 | 0.64 | 0.19 |
| LR | 0.65 | 0.56 | 0.61 | 0.65 | 0.21 | 0.63 | 0.56 | 0.59 | 0.62 | 0.18 |
| GB | 0.69 | 0.54 | 0.61 | 0.66 | 0.23 | 0.67 | 0.52 | 0.60 | 0.64 | 0.20 |
| LGBM | 0.66 | 0.56 | 0.61 | 0.65 | 0.22 | 0.63 | 0.54 | 0.59 | 0.62 | 0.18 |

### Window 11

| Model | Training Dataset | | | | | Validation dataset | | | | |
|---|---|---|---|---|---|---|---|---|---|---|
| | Sens | Spec | Acc | AUROC | MCC | Sens | Spec | Acc | AUROC | MCC |
| NB | 0.70 | 0.50 | 0.60 | 0.64 | 0.21 | 0.64 | 0.54 | 0.59 | 0.62 | 0.18 |
| RF | 0.68 | 0.56 | 0.62 | 0.67 | 0.24 | 0.67 | 0.54 | 0.60 | 0.64 | 0.21 |
| LR | 0.64 | 0.57 | 0.60 | 0.65 | 0.21 | 0.62 | 0.54 | 0.58 | 0.62 | 0.16 |
| GB | 0.69 | 0.54 | 0.62 | 0.66 | 0.23 | 0.68 | 0.53 | 0.60 | 0.63 | 0.21 |
| LGBM | 0.66 | 0.58 | 0.62 | 0.66 | 0.24 | 0.63 | 0.55 | 0.59 | 0.63 | 0.18 |

### Window 13

| Model | Training Dataset | | | | | Validation dataset | | | | |
|---|---|---|---|---|---|---|---|---|---|---|
| | Sens | Spec | Acc | AUROC | MCC | Sens | Spec | Acc | AUROC | MCC |
| NB | 0.70 | 0.50 | 0.60 | 0.64 | 0.20 | 0.62 | 0.53 | 0.58 | 0.61 | 0.15 |
| RF | 0.69 | 0.56 | 0.62 | 0.67 | 0.25 | 0.65 | 0.54 | 0.60 | 0.64 | 0.19 |
| LR | 0.63 | 0.56 | 0.60 | 0.64 | 0.19 | 0.60 | 0.53 | 0.56 | 0.60 | 0.13 |
| GB | 0.68 | 0.54 | 0.61 | 0.66 | 0.23 | 0.67 | 0.52 | 0.60 | 0.62 | 0.19 |
| LGBM | 0.66 | 0.58 | 0.62 | 0.66 | 0.24 | 0.61 | 0.54 | 0.58 | 0.61 | 0.15 |

| | Window 15 | | | | | | | | | |
|---|---|---|---|---|---|---|---|---|---|---|
| **Model** | **Training Dataset** | | | | | **Validation dataset** | | | | |
| | Sens | Spec | Acc | AUROC | MCC | Sens | Spec | Acc | AUROC | MCC |
| **NB** | 0.69 | 0.50 | 0.60 | 0.64 | 0.20 | 0.64 | 0.53 | 0.58 | 0.61 | 0.17 |
| **RF** | 0.68 | 0.55 | 0.61 | 0.67 | 0.23 | 0.65 | 0.53 | 0.59 | 0.64 | 0.18 |
| **LR** | 0.63 | 0.57 | 0.60 | 0.64 | 0.20 | 0.61 | 0.52 | 0.56 | 0.60 | 0.13 |
| **GB** | 0.69 | 0.55 | 0.62 | 0.66 | 0.24 | 0.67 | 0.52 | 0.59 | 0.63 | 0.19 |
| **LGBM** | 0.64 | 0.58 | 0.61 | 0.65 | 0.22 | 0.62 | 0.56 | 0.59 | 0.63 | 0.18 |

| | Window 17 | | | | | | | | | |
|---|---|---|---|---|---|---|---|---|---|---|
| **Model** | **Training Dataset** | | | | | **Validation dataset** | | | | |
| | Sens | Spec | Acc | AUROC | MCC | Sens | Spec | Acc | AUROC | MCC |
| **NB** | 0.69 | 0.51 | 0.60 | 0.64 | 0.20 | 0.61 | 0.54 | 0.58 | 0.60 | 0.15 |
| **RF** | 0.68 | 0.56 | 0.62 | 0.67 | 0.24 | 0.63 | 0.53 | 0.58 | 0.63 | 0.16 |
| **LR** | 0.63 | 0.57 | 0.60 | 0.65 | 0.20 | 0.59 | 0.52 | 0.56 | 0.59 | 0.12 |
| **GB** | 0.69 | 0.55 | 0.62 | 0.66 | 0.24 | 0.65 | 0.52 | 0.58 | 0.62 | 0.17 |
| **LGBM** | 0.65 | 0.58 | 0.61 | 0.66 | 0.23 | 0.59 | 0.54 | 0.56 | 0.60 | 0.12 |

| | Window 19 | | | | | | | | | |
|---|---|---|---|---|---|---|---|---|---|---|
| **Model** | **Training Dataset** | | | | | **Validation dataset** | | | | |
| | Sens | Spec | Acc | AUROC | MCC | Sens | Spec | Acc | AUROC | MCC |
| **NB** | 0.69 | 0.50 | 0.59 | 0.63 | 0.19 | 0.61 | 0.54 | 0.57 | 0.60 | 0.15 |
| **RF** | 0.66 | 0.56 | 0.61 | 0.66 | 0.22 | 0.62 | 0.54 | 0.58 | 0.62 | 0.16 |
| **LR** | 0.62 | 0.57 | 0.60 | 0.64 | 0.19 | 0.59 | 0.53 | 0.56 | 0.59 | 0.12 |
| **GB** | 0.67 | 0.55 | 0.61 | 0.65 | 0.22 | 0.65 | 0.52 | 0.59 | 0.62 | 0.18 |
| **LGBM** | 0.64 | 0.57 | 0.60 | 0.65 | 0.21 | 0.61 | 0.54 | 0.58 | 0.62 | 0.15 |

| | Window 21 | | | | | | | | | |
|---|---|---|---|---|---|---|---|---|---|---|
| **Model** | **Training Dataset** | | | | | **Validation dataset** | | | | |
| | Sens | Spec | Acc | AUROC | MCC | Sens | Spec | Acc | AUROC | MCC |
| **NB** | 0.69 | 0.51 | 0.60 | 0.64 | 0.21 | 0.60 | 0.54 | 0.57 | 0.60 | 0.15 |
| **RF** | 0.68 | 0.56 | 0.62 | 0.67 | 0.24 | 0.64 | 0.55 | 0.59 | 0.63 | 0.18 |
| **LR** | 0.62 | 0.58 | 0.60 | 0.64 | 0.20 | 0.58 | 0.54 | 0.56 | 0.59 | 0.12 |
| **GB** | 0.69 | 0.55 | 0.62 | 0.66 | 0.24 | 0.65 | 0.51 | 0.58 | 0.62 | 0.16 |
| **LGBM** | 0.64 | 0.58 | 0.61 | 0.65 | 0.23 | 0.59 | 0.55 | 0.57 | 0.61 | 0.14 |

| | Window 23 | | | | | | | | | |
|---|---|---|---|---|---|---|---|---|---|---|
| **Model** | **Training Dataset** | | | | | **Validation dataset** | | | | |
| | Sens | Spec | Acc | AUROC | MCC | Sens | Spec | Acc | AUROC | MCC |
| **NB** | 0.68 | 0.50 | 0.59 | 0.63 | 0.19 | 0.59 | 0.55 | 0.57 | 0.60 | 0.14 |
| **RF** | 0.67 | 0.54 | 0.61 | 0.66 | 0.21 | 0.62 | 0.55 | 0.58 | 0.63 | 0.17 |
| **LR** | 0.62 | 0.58 | 0.60 | 0.64 | 0.19 | 0.56 | 0.54 | 0.55 | 0.59 | 0.11 |
| GB | 0.68 | 0.55 | 0.62 | 0.66 | 0.24 | 0.64 | 0.54 | 0.59 | 0.63 | 0.18 |
| LGBM | 0.65 | 0.58 | 0.62 | 0.65 | 0.23 | 0.58 | 0.56 | 0.57 | 0.62 | 0.15 |

| | Window 25 | | | | | | | | | |
|---|---|---|---|---|---|---|---|---|---|---|
| **Model** | **Training Dataset** | | | | | **Validation dataset** | | | | |
| | Sens | Spec | Acc | AUROC | MCC | Sens | Spec | Acc | AUROC | MCC |
| **NB** | 0.68 | 0.50 | 0.59 | 0.62 | 0.18 | 0.59 | 0.54 | 0.56 | 0.59 | 0.13 |
| **RF** | 0.67 | 0.56 | 0.61 | 0.66 | 0.23 | 0.64 | 0.54 | 0.59 | 0.64 | 0.19 |
| **LR** | 0.61 | 0.57 | 0.59 | 0.63 | 0.18 | 0.57 | 0.54 | 0.56 | 0.58 | 0.11 |
| **GB** | 0.67 | 0.54 | 0.60 | 0.65 | 0.21 | 0.65 | 0.53 | 0.59 | 0.63 | 0.18 |
| **LGBM** | 0.63 | 0.57 | 0.60 | 0.65 | 0.21 | 0.60 | 0.56 | 0.58 | 0.63 | 0.17 |

| | Window 27 | | | | | | | | | |
|---|---|---|---|---|---|---|---|---|---|---|
| **Model** | **Training Dataset** | | | | | **Validation dataset** | | | | |
| | Sens | Spec | Acc | AUROC | MCC | Sens | Spec | Acc | AUROC | MCC |
| **NB** | 0.68 | 0.49 | 0.58 | 0.62 | 0.17 | 0.59 | 0.54 | 0.56 | 0.59 | 0.13 |
| **RF** | 0.67 | 0.55 | 0.61 | 0.66 | 0.22 | 0.65 | 0.54 | 0.59 | 0.64 | 0.19 |
| **LR** | 0.61 | 0.57 | 0.59 | 0.63 | 0.18 | 0.59 | 0.52 | 0.56 | 0.59 | 0.11 |
| **GB** | 0.67 | 0.54 | 0.60 | 0.65 | 0.21 | 0.64 | 0.53 | 0.59 | 0.63 | 0.18 |
| **LGBM** | 0.63 | 0.57 | 0.60 | 0.64 | 0.20 | 0.59 | 0.56 | 0.57 | 0.61 | 0.14 |

| | Window 29 | | | | | | | | | |
|---|---|---|---|---|---|---|---|---|---|---|
| **Model** | **Training Dataset** | | | | | **Validation dataset** | | | | |
| | Sens | Spec | Acc | AUROC | MCC | Sens | Spec | Acc | AUROC | MCC |
| **NB** | 0.67 | 0.50 | 0.58 | 0.62 | 0.17 | 0.58 | 0.55 | 0.56 | 0.59 | 0.13 |
| **RF** | 0.67 | 0.54 | 0.60 | 0.65 | 0.21 | 0.64 | 0.54 | 0.59 | 0.62 | 0.18 |
| **LR** | 0.60 | 0.57 | 0.59 | 0.62 | 0.17 | 0.59 | 0.52 | 0.55 | 0.58 | 0.11 |
| **GB** | 0.68 | 0.55 | 0.61 | 0.65 | 0.22 | 0.65 | 0.53 | 0.59 | 0.62 | 0.18 |
| **LGBM** | 0.65 | 0.58 | 0.62 | 0.65 | 0.23 | 0.61 | 0.56 | 0.59 | 0.62 | 0.18 |

| | Window 31 | | | | | | | | | |
|---|---|---|---|---|---|---|---|---|---|---|
| **Model** | **Training Dataset** | | | | | **Validation dataset** | | | | |
| | Sens | Spec | Acc | AUROC | MCC | Sens | Spec | Acc | AUROC | MCC |
| **NB** | 0.67 | 0.50 | 0.58 | 0.62 | 0.17 | 0.60 | 0.54 | 0.57 | 0.58 | 0.13 |
| **RF** | 0.67 | 0.56 | 0.61 | 0.66 | 0.23 | 0.64 | 0.53 | 0.59 | 0.63 | 0.17 |
| **LR** | 0.59 | 0.57 | 0.58 | 0.62 | 0.17 | 0.56 | 0.53 | 0.54 | 0.58 | 0.09 |
| **GB** | 0.66 | 0.54 | 0.60 | 0.64 | 0.21 | 0.64 | 0.54 | 0.59 | 0.63 | 0.18 |
| **LGBM** | 0.64 | 0.58 | 0.61 | 0.65 | 0.22 | 0.62 | 0.58 | 0.60 | 0.63 | 0.20 |

## Supplementary Table S3: The complete result on different windows using PSSM profile + Binary profiles

LR, Logistic Regression; RF, Random Forest; NB, Naive Bayes; LGBM, Light Gradient Boosting Machine; GB, Gradient Boost; Sens, Sensitivity; Spec, Specificity; Acc, Accuracy; AUROC, Area Under the Receiver Operating Characteristic; MCC, Matthews Correlation Coefficient

### Window 7

| Model | Training Dataset | | | | | Validation dataset | | | | |
|---|---|---|---|---|---|---|---|---|---|---|
| | Sens | Spec | Acc | AUROC | MCC | Sens | Spec | Acc | AUROC | MCC |
| RF | 0.69 | 0.56 | 0.62 | 0.67 | 0.25 | 0.65 | 0.54 | 0.60 | 0.64 | 0.19 |
| LGBM | 0.66 | 0.57 | 0.61 | 0.66 | 0.23 | 0.62 | 0.56 | 0.59 | 0.63 | 0.18 |
| GB | 0.69 | 0.54 | 0.62 | 0.66 | 0.24 | 0.69 | 0.52 | 0.60 | 0.64 | 0.21 |
| NB | 0.68 | 0.53 | 0.61 | 0.65 | 0.22 | 0.64 | 0.55 | 0.59 | 0.63 | 0.19 |
| LR | 0.63 | 0.57 | 0.60 | 0.65 | 0.21 | 0.63 | 0.56 | 0.59 | 0.63 | 0.19 |

### Window 9

| Model | Training Dataset | | | | | Validation dataset | | | | |
|---|---|---|---|---|---|---|---|---|---|---|
| | Sens | Spec | Acc | AUROC | MCC | Sens | Spec | Acc | AUROC | MCC |
| RF | 0.67 | 0.56 | 0.62 | 0.66 | 0.23 | 0.63 | 0.54 | 0.59 | 0.63 | 0.17 |
| LGBM | 0.66 | 0.57 | 0.61 | 0.65 | 0.23 | 0.60 | 0.55 | 0.58 | 0.62 | 0.16 |
| GB | 0.69 | 0.55 | 0.62 | 0.66 | 0.24 | 0.68 | 0.52 | 0.60 | 0.63 | 0.20 |
| NB | 0.68 | 0.54 | 0.61 | 0.64 | 0.21 | 0.62 | 0.56 | 0.59 | 0.63 | 0.19 |
| LR | 0.62 | 0.58 | 0.60 | 0.64 | 0.20 | 0.62 | 0.57 | 0.59 | 0.63 | 0.18 |

### Window 11

| Model | Training Dataset | | | | | Validation dataset | | | | |
|---|---|---|---|---|---|---|---|---|---|---|
| | Sens | Spec | Acc | AUROC | MCC | Sens | Spec | Acc | AUROC | MCC |
| RF | 0.68 | 0.56 | 0.62 | 0.67 | 0.25 | 0.65 | 0.55 | 0.60 | 0.64 | 0.20 |
| LGBM | 0.66 | 0.57 | 0.61 | 0.66 | 0.23 | 0.63 | 0.55 | 0.59 | 0.64 | 0.19 |
| GB | 0.69 | 0.55 | 0.62 | 0.66 | 0.24 | 0.66 | 0.52 | 0.59 | 0.63 | 0.19 |
| NB | 0.67 | 0.54 | 0.60 | 0.64 | 0.21 | 0.62 | 0.56 | 0.59 | 0.62 | 0.18 |
| LR | 0.62 | 0.58 | 0.60 | 0.64 | 0.19 | 0.62 | 0.56 | 0.59 | 0.62 | 0.17 |

### Window 13

| Model | Training Dataset | | | | | Validation dataset | | | | |
|---|---|---|---|---|---|---|---|---|---|---|
| | Sens | Spec | Acc | AUROC | MCC | Sens | Spec | Acc | AUROC | MCC |
| RF | 0.68 | 0.56 | 0.62 | 0.66 | 0.24 | 0.63 | 0.53 | 0.58 | 0.63 | 0.16 |
| LGBM | 0.65 | 0.58 | 0.61 | 0.65 | 0.23 | 0.61 | 0.55 | 0.58 | 0.61 | 0.17 |
| GB | 0.69 | 0.54 | 0.61 | 0.66 | 0.23 | 0.67 | 0.53 | 0.60 | 0.63 | 0.20 |
| NB | 0.66 | 0.53 | 0.59 | 0.64 | 0.19 | 0.61 | 0.55 | 0.58 | 0.61 | 0.16 |
| LR | 0.61 | 0.59 | 0.60 | 0.63 | 0.20 | 0.61 | 0.55 | 0.58 | 0.60 | 0.16 |

### Window 15

| Model | Training Dataset | | | | | Validation dataset | | | | |
|---|---|---|---|---|---|---|---|---|---|---|
| | Sens | Spec | Acc | AUROC | MCC | Sens | Spec | Acc | AUROC | MCC |
| RF | 0.67 | 0.56 | 0.62 | 0.67 | 0.23 | 0.66 | 0.53 | 0.59 | 0.63 | 0.19 |
| LGBM | 0.65 | 0.58 | 0.61 | 0.66 | 0.22 | 0.62 | 0.55 | 0.59 | 0.63 | 0.17 |
| GB | 0.69 | 0.55 | 0.62 | 0.66 | 0.25 | 0.68 | 0.52 | 0.60 | 0.64 | 0.21 |
| NB | 0.65 | 0.54 | 0.59 | 0.63 | 0.19 | 0.62 | 0.55 | 0.59 | 0.62 | 0.17 |
| LR | 0.61 | 0.58 | 0.59 | 0.63 | 0.19 | 0.62 | 0.54 | 0.58 | 0.60 | 0.16 |

| Window 17 | | | | | | | | | | |
|---|---|---|---|---|---|---|---|---|---|---|
| Model | Training Dataset | | | | | Validation dataset | | | | |
| | Sens | Spec | Acc | AUROC | MCC | Sens | Spec | Acc | AUROC | MCC |
| RF | 0.67 | 0.55 | 0.61 | 0.66 | 0.22 | 0.64 | 0.54 | 0.59 | 0.63 | 0.18 |
| LGBM | 0.65 | 0.57 | 0.61 | 0.65 | 0.21 | 0.61 | 0.56 | 0.58 | 0.62 | 0.16 |
| GB | 0.69 | 0.55 | 0.62 | 0.66 | 0.24 | 0.64 | 0.52 | 0.58 | 0.62 | 0.17 |
| NB | 0.65 | 0.54 | 0.60 | 0.63 | 0.19 | 0.60 | 0.55 | 0.58 | 0.60 | 0.15 |
| LR | 0.60 | 0.58 | 0.59 | 0.63 | 0.18 | 0.58 | 0.53 | 0.56 | 0.59 | 0.11 |

| Window 19 | | | | | | | | | | |
|---|---|---|---|---|---|---|---|---|---|---|
| Model | Training Dataset | | | | | Validation dataset | | | | |
| | Sens | Spec | Acc | AUROC | MCC | Sens | Spec | Acc | AUROC | MCC |
| RF | 0.67 | 0.55 | 0.61 | 0.66 | 0.22 | 0.62 | 0.54 | 0.58 | 0.62 | 0.16 |
| LGBM | 0.65 | 0.57 | 0.61 | 0.65 | 0.22 | 0.59 | 0.55 | 0.57 | 0.62 | 0.14 |
| GB | 0.68 | 0.54 | 0.61 | 0.65 | 0.22 | 0.65 | 0.52 | 0.58 | 0.62 | 0.17 |
| NB | 0.63 | 0.55 | 0.59 | 0.62 | 0.18 | 0.58 | 0.56 | 0.57 | 0.60 | 0.15 |
| LR | 0.59 | 0.58 | 0.58 | 0.63 | 0.17 | 0.57 | 0.54 | 0.56 | 0.59 | 0.11 |

| Window 21 | | | | | | | | | | |
|---|---|---|---|---|---|---|---|---|---|---|
| Model | Training Dataset | | | | | Validation dataset | | | | |
| | Sens | Spec | Acc | AUROC | MCC | Sens | Spec | Acc | AUROC | MCC |
| RF | 0.68 | 0.56 | 0.62 | 0.66 | 0.24 | 0.64 | 0.54 | 0.59 | 0.64 | 0.18 |
| LGBM | 0.65 | 0.58 | 0.61 | 0.66 | 0.23 | 0.58 | 0.55 | 0.56 | 0.61 | 0.13 |
| GB | 0.69 | 0.54 | 0.61 | 0.65 | 0.23 | 0.65 | 0.50 | 0.57 | 0.62 | 0.15 |
| NB | 0.63 | 0.55 | 0.59 | 0.63 | 0.18 | 0.57 | 0.55 | 0.56 | 0.60 | 0.12 |
| LR | 0.59 | 0.57 | 0.58 | 0.62 | 0.16 | 0.57 | 0.54 | 0.56 | 0.59 | 0.11 |

| Window 23 | | | | | | | | | | |
|---|---|---|---|---|---|---|---|---|---|---|
| Model | Training Dataset | | | | | Validation dataset | | | | |
| | Sens | Spec | Acc | AUROC | MCC | Sens | Spec | Acc | AUROC | MCC |
| RF | 0.66 | 0.56 | 0.61 | 0.66 | 0.23 | 0.62 | 0.55 | 0.58 | 0.63 | 0.16 |
| LGBM | 0.65 | 0.58 | 0.61 | 0.66 | 0.22 | 0.60 | 0.56 | 0.58 | 0.62 | 0.17 |
| GB | 0.69 | 0.55 | 0.62 | 0.66 | 0.24 | 0.64 | 0.53 | 0.58 | 0.63 | 0.17 |
| NB | 0.63 | 0.55 | 0.59 | 0.63 | 0.18 | 0.57 | 0.55 | 0.56 | 0.60 | 0.12 |
| LR | 0.59 | 0.57 | 0.58 | 0.62 | 0.16 | 0.55 | 0.54 | 0.54 | 0.58 | 0.09 |

| Window 25 | | | | | | | | | | |
|---|---|---|---|---|---|---|---|---|---|---|
| Model | Training Dataset | | | | | Validation dataset | | | | |
| | Sens | Spec | Acc | AUROC | MCC | Sens | Spec | Acc | AUROC | MCC |
| RF | 0.67 | 0.54 | 0.61 | 0.66 | 0.21 | 0.65 | 0.54 | 0.59 | 0.64 | 0.19 |
| LGBM | 0.64 | 0.56 | 0.60 | 0.65 | 0.20 | 0.60 | 0.57 | 0.59 | 0.62 | 0.17 |
| GB | 0.68 | 0.53 | 0.61 | 0.65 | 0.22 | 0.66 | 0.51 | 0.59 | 0.63 | 0.18 |
| NB | 0.62 | 0.55 | 0.58 | 0.62 | 0.17 | 0.57 | 0.55 | 0.56 | 0.59 | 0.12 |
| LR | 0.58 | 0.56 | 0.57 | 0.61 | 0.15 | 0.56 | 0.54 | 0.55 | 0.57 | 0.10 |

| Window 27 | | | | | | | | | | |
|---|---|---|---|---|---|---|---|---|---|---|
| Model | Training Dataset | | | | | Validation dataset | | | | |
| | Sens | Spec | Acc | AUROC | MCC | Sens | Spec | Acc | AUROC | MCC |
| RF | 0.67 | 0.55 | 0.61 | 0.65 | 0.22 | 0.63 | 0.53 | 0.58 | 0.63 | 0.17 |
| LGBM | 0.64 | 0.58 | 0.61 | 0.65 | 0.22 | 0.62 | 0.57 | 0.59 | 0.62 | 0.19 |
| NB | 0.62 | 0.54 | 0.58 | 0.62 | 0.16 | 0.58 | 0.55 | 0.56 | 0.59 | 0.13 |
| LR | 0.59 | 0.57 | 0.58 | 0.61 | 0.16 | 0.57 | 0.54 | 0.55 | 0.58 | 0.11 |
| GB | 0.68 | 0.53 | 0.61 | 0.65 | 0.21 | 0.66 | 0.54 | 0.60 | 0.63 | 0.20 |

| Window 29 | | | | | | | | | | |
|---|---|---|---|---|---|---|---|---|---|---|
| Model | Training Dataset | | | | | Validation dataset | | | | |
| | Sens | Spec | Acc | AUROC | MCC | Sens | Spec | Acc | AUROC | MCC |
| RF | 0.67 | 0.54 | 0.61 | 0.65 | 0.21 | 0.63 | 0.54 | 0.58 | 0.63 | 0.17 |
| LGBM | 0.64 | 0.58 | 0.61 | 0.66 | 0.22 | 0.60 | 0.55 | 0.57 | 0.62 | 0.15 |
| NB | 0.61 | 0.55 | 0.58 | 0.61 | 0.16 | 0.58 | 0.55 | 0.56 | 0.59 | 0.13 |
| LR | 0.59 | 0.56 | 0.58 | 0.61 | 0.15 | 0.57 | 0.54 | 0.56 | 0.58 | 0.11 |
| GB | 0.68 | 0.54 | 0.61 | 0.65 | 0.22 | 0.65 | 0.53 | 0.59 | 0.62 | 0.18 |

| Window 31 | | | | | | | | | | |
|---|---|---|---|---|---|---|---|---|---|---|
| Model | Training Dataset | | | | | Validation dataset | | | | |
| | Sens | Spec | Acc | AUROC | MCC | Sens | Spec | Acc | AUROC | MCC |
| RF | 0.66 | 0.55 | 0.61 | 0.65 | 0.22 | 0.62 | 0.54 | 0.58 | 0.62 | 0.16 |
| LGBM | 0.64 | 0.58 | 0.61 | 0.65 | 0.22 | 0.60 | 0.57 | 0.58 | 0.63 | 0.16 |
| NB | 0.61 | 0.55 | 0.58 | 0.61 | 0.16 | 0.57 | 0.56 | 0.56 | 0.59 | 0.13 |
| LR | 0.58 | 0.57 | 0.57 | 0.60 | 0.15 | 0.56 | 0.53 | 0.55 | 0.57 | 0.09 |
| GB | 0.67 | 0.54 | 0.61 | 0.64 | 0.21 | 0.66 | 0.53 | 0.60 | 0.63 | 0.20 |

## Supplementary Table S4: The complete result on different windows using PSSM profile + RSA features

LR, Logistic Regression; RF, Random Forest; NB, Naive Bayes; LGBM, Light Gradient Boosting Machine; GB, Gradient Boost; Sens, Sensitivity; Spec, Specificity; Acc, Accuracy; AUROC, Area Under the Receiver Operating Characteristic; MCC, Matthews Correlation Coefficient

### Window 7

| Model | Training Dataset | | | | | Validation dataset | | | | |
|---|---|---|---|---|---|---|---|---|---|---|
| | Sens | Spec | Acc | AUROC | MCC | Sens | Spec | Acc | AUROC | MCC |
| **RF** | 0.68 | 0.57 | 0.63 | 0.67 | 0.25 | 0.67 | 0.54 | 0.60 | 0.64 | 0.21 |
| **LGBM** | 0.67 | 0.57 | 0.62 | 0.66 | 0.24 | 0.64 | 0.54 | 0.59 | 0.63 | 0.19 |
| **NB** | 0.73 | 0.50 | 0.61 | 0.65 | 0.23 | 0.68 | 0.52 | 0.60 | 0.63 | 0.20 |
| **LR** | 0.66 | 0.58 | 0.62 | 0.67 | 0.25 | 0.63 | 0.56 | 0.60 | 0.63 | 0.19 |
| **GB** | 0.71 | 0.55 | 0.63 | 0.67 | 0.26 | 0.69 | 0.51 | 0.60 | 0.64 | 0.20 |

### Window 9

| Model | Training Dataset | | | | | Validation dataset | | | | |
|---|---|---|---|---|---|---|---|---|---|---|
| | Sens | Spec | Acc | AUROC | MCC | Sens | Spec | Acc | AUROC | MCC |
| **RF** | 0.68 | 0.56 | 0.62 | 0.67 | 0.24 | 0.66 | 0.55 | 0.60 | 0.65 | 0.20 |
| **LGBM** | 0.67 | 0.56 | 0.62 | 0.66 | 0.24 | 0.63 | 0.54 | 0.59 | 0.64 | 0.17 |
| **NB** | 0.70 | 0.52 | 0.61 | 0.65 | 0.22 | 0.65 | 0.54 | 0.59 | 0.63 | 0.19 |
| **LR** | 0.65 | 0.59 | 0.62 | 0.66 | 0.25 | 0.61 | 0.57 | 0.59 | 0.63 | 0.19 |
| **GB** | 0.70 | 0.55 | 0.62 | 0.67 | 0.25 | 0.67 | 0.53 | 0.60 | 0.64 | 0.20 |

### Window 11

| Model | Training Dataset | | | | | Validation dataset | | | | |
|---|---|---|---|---|---|---|---|---|---|---|
| | Sens | Spec | Acc | AUROC | MCC | Sens | Spec | Acc | AUROC | MCC |
| **RF** | 0.69 | 0.57 | 0.63 | 0.67 | 0.26 | 0.65 | 0.52 | 0.58 | 0.63 | 0.17 |
| **LGBM** | 0.67 | 0.58 | 0.63 | 0.67 | 0.26 | 0.63 | 0.55 | 0.59 | 0.63 | 0.18 |
| **NB** | 0.74 | 0.48 | 0.61 | 0.65 | 0.23 | 0.68 | 0.52 | 0.60 | 0.63 | 0.20 |
| **LR** | 0.65 | 0.59 | 0.62 | 0.66 | 0.24 | 0.61 | 0.56 | 0.58 | 0.62 | 0.17 |
| **GB** | 0.70 | 0.55 | 0.63 | 0.67 | 0.26 | 0.68 | 0.53 | 0.61 | 0.64 | 0.22 |

### Window 13

| Model | Training Dataset | | | | | Validation dataset | | | | |
|---|---|---|---|---|---|---|---|---|---|---|
| | Sens | Spec | Acc | AUROC | MCC | Sens | Spec | Acc | AUROC | MCC |
| **RF** | 0.68 | 0.56 | 0.62 | 0.67 | 0.24 | 0.64 | 0.54 | 0.59 | 0.63 | 0.18 |
| **LGBM** | 0.66 | 0.57 | 0.62 | 0.66 | 0.23 | 0.62 | 0.55 | 0.59 | 0.62 | 0.17 |
| **NB** | 0.74 | 0.47 | 0.60 | 0.64 | 0.22 | 0.66 | 0.51 | 0.59 | 0.61 | 0.18 |
| **LR** | 0.64 | 0.59 | 0.61 | 0.66 | 0.23 | 0.59 | 0.54 | 0.57 | 0.61 | 0.13 |
| **GB** | 0.70 | 0.56 | 0.63 | 0.67 | 0.26 | 0.66 | 0.52 | 0.59 | 0.61 | 0.18 |

### Window 15

| Model | Training Dataset | | | | | Validation dataset | | | | |
|---|---|---|---|---|---|---|---|---|---|---|
| | Sens | Spec | Acc | AUROC | MCC | Sens | Spec | Acc | AUROC | MCC |
| RF | 0.69 | 0.56 | 0.63 | 0.67 | 0.26 | 0.66 | 0.56 | 0.61 | 0.64 | 0.22 |
| LGBM | 0.66 | 0.58 | 0.62 | 0.66 | 0.24 | 0.64 | 0.56 | 0.60 | 0.64 | 0.20 |
| NB | 0.72 | 0.49 | 0.60 | 0.64 | 0.21 | 0.68 | 0.51 | 0.60 | 0.62 | 0.19 |
| LR | 0.64 | 0.59 | 0.61 | 0.66 | 0.23 | 0.61 | 0.54 | 0.57 | 0.61 | 0.14 |
| GB | 0.69 | 0.56 | 0.62 | 0.67 | 0.25 | 0.68 | 0.52 | 0.60 | 0.64 | 0.20 |

### Window 17

| Model | Training Dataset | | | | | Validation dataset | | | | |
|---|---|---|---|---|---|---|---|---|---|---|
| | Sens | Spec | Acc | AUROC | MCC | Sens | Spec | Acc | AUROC | MCC |
| RF | 0.67 | 0.56 | 0.61 | 0.66 | 0.23 | 0.64 | 0.54 | 0.59 | 0.64 | 0.18 |
| LGBM | 0.65 | 0.58 | 0.62 | 0.66 | 0.23 | 0.63 | 0.55 | 0.59 | 0.62 | 0.17 |
| NB | 0.73 | 0.47 | 0.60 | 0.64 | 0.21 | 0.65 | 0.53 | 0.59 | 0.61 | 0.18 |
| LR | 0.64 | 0.59 | 0.62 | 0.66 | 0.23 | 0.58 | 0.55 | 0.57 | 0.60 | 0.14 |
| GB | 0.70 | 0.55 | 0.63 | 0.67 | 0.25 | 0.65 | 0.53 | 0.59 | 0.62 | 0.19 |

### Window 19

| Model | Training Dataset | | | | | Validation dataset | | | | |
|---|---|---|---|---|---|---|---|---|---|---|
| | Sens | Spec | Acc | AUROC | MCC | Sens | Spec | Acc | AUROC | MCC |
| RF | 0.67 | 0.56 | 0.61 | 0.66 | 0.23 | 0.64 | 0.53 | 0.59 | 0.63 | 0.17 |
| LGBM | 0.65 | 0.56 | 0.61 | 0.65 | 0.21 | 0.64 | 0.55 | 0.60 | 0.63 | 0.19 |
| NB | 0.67 | 0.52 | 0.59 | 0.63 | 0.19 | 0.59 | 0.56 | 0.57 | 0.60 | 0.15 |
| LR | 0.63 | 0.59 | 0.61 | 0.66 | 0.22 | 0.60 | 0.55 | 0.58 | 0.61 | 0.15 |
| GB | 0.69 | 0.54 | 0.62 | 0.66 | 0.24 | 0.65 | 0.53 | 0.59 | 0.62 | 0.17 |

### Window 21

| Model | Training Dataset | | | | | Validation dataset | | | | |
|---|---|---|---|---|---|---|---|---|---|---|
| | Sens | Spec | Acc | AUROC | MCC | Sens | Spec | Acc | AUROC | MCC |
| RF | 0.68 | 0.55 | 0.62 | 0.67 | 0.24 | 0.64 | 0.54 | 0.59 | 0.64 | 0.18 |
| LGBM | 0.65 | 0.57 | 0.61 | 0.66 | 0.23 | 0.60 | 0.56 | 0.58 | 0.63 | 0.17 |
| NB | 0.72 | 0.48 | 0.60 | 0.64 | 0.21 | 0.62 | 0.54 | 0.58 | 0.60 | 0.16 |
| LR | 0.63 | 0.60 | 0.61 | 0.66 | 0.23 | 0.60 | 0.55 | 0.57 | 0.61 | 0.15 |
| GB | 0.69 | 0.56 | 0.62 | 0.67 | 0.25 | 0.65 | 0.52 | 0.59 | 0.62 | 0.17 |

### Window 23

| Model | Training Dataset | | | | | Validation dataset | | | | |
|---|---|---|---|---|---|---|---|---|---|---|
| | Sens | Spec | Acc | AUROC | MCC | Sens | Spec | Acc | AUROC | MCC |
| RF | 0.68 | 0.56 | 0.62 | 0.67 | 0.24 | 0.63 | 0.54 | 0.59 | 0.63 | 0.18 |
| LGBM | 0.66 | 0.57 | 0.61 | 0.66 | 0.22 | 0.60 | 0.55 | 0.57 | 0.62 | 0.15 |
| NB | 0.68 | 0.51 | 0.59 | 0.63 | 0.19 | 0.59 | 0.55 | 0.57 | 0.60 | 0.15 |
| LR | 0.63 | 0.60 | 0.61 | 0.66 | 0.23 | 0.59 | 0.58 | 0.58 | 0.61 | 0.17 |

| | Sens | Spec | Acc | AUROC | MCC | Sens | Spec | Acc | AUROC | MCC |
|---|---|---|---|---|---|---|---|---|---|---|
| GB | 0.69 | 0.55 | 0.62 | 0.67 | 0.24 | 0.64 | 0.53 | 0.59 | 0.63 | 0.17 |

| Window 25 ||||||||||
|---|---|---|---|---|---|---|---|---|---|---|
| Model | Training Dataset ||||| Validation dataset |||||
| | Sens | Spec | Acc | AUROC | MCC | Sens | Spec | Acc | AUROC | MCC |
| RF | 0.67 | 0.54 | 0.61 | 0.65 | 0.21 | 0.65 | 0.54 | 0.60 | 0.64 | 0.19 |
| LGBM | 0.64 | 0.57 | 0.61 | 0.66 | 0.22 | 0.61 | 0.55 | 0.58 | 0.63 | 0.16 |
| NB | 0.65 | 0.54 | 0.60 | 0.62 | 0.19 | 0.56 | 0.57 | 0.57 | 0.58 | 0.13 |
| LR | 0.62 | 0.59 | 0.61 | 0.65 | 0.21 | 0.59 | 0.56 | 0.57 | 0.60 | 0.15 |
| GB | 0.68 | 0.54 | 0.61 | 0.66 | 0.23 | 0.66 | 0.53 | 0.60 | 0.64 | 0.20 |

| Window 27 ||||||||||
|---|---|---|---|---|---|---|---|---|---|---|
| Model | Training Dataset ||||| Validation dataset |||||
| | Sens | Spec | Acc | AUROC | MCC | Sens | Spec | Acc | AUROC | MCC |
| RF | 0.68 | 0.55 | 0.61 | 0.66 | 0.23 | 0.65 | 0.53 | 0.59 | 0.63 | 0.18 |
| LGBM | 0.64 | 0.59 | 0.61 | 0.66 | 0.23 | 0.61 | 0.56 | 0.58 | 0.63 | 0.16 |
| NB | 0.66 | 0.52 | 0.59 | 0.62 | 0.18 | 0.57 | 0.57 | 0.57 | 0.59 | 0.14 |
| LR | 0.63 | 0.59 | 0.61 | 0.65 | 0.22 | 0.60 | 0.56 | 0.58 | 0.61 | 0.16 |
| GB | 0.69 | 0.55 | 0.62 | 0.66 | 0.24 | 0.67 | 0.53 | 0.60 | 0.64 | 0.20 |

| Window 29 ||||||||||
|---|---|---|---|---|---|---|---|---|---|---|
| Model | Training Dataset ||||| Validation dataset |||||
| | Sens | Spec | Acc | AUROC | MCC | Sens | Spec | Acc | AUROC | MCC |
| RF | 0.68 | 0.56 | 0.62 | 0.66 | 0.24 | 0.63 | 0.53 | 0.58 | 0.62 | 0.16 |
| LGBM | 0.65 | 0.58 | 0.61 | 0.66 | 0.23 | 0.59 | 0.57 | 0.58 | 0.63 | 0.16 |
| NB | 0.67 | 0.51 | 0.59 | 0.62 | 0.18 | 0.58 | 0.56 | 0.57 | 0.59 | 0.14 |
| LR | 0.62 | 0.59 | 0.61 | 0.65 | 0.21 | 0.58 | 0.55 | 0.56 | 0.60 | 0.13 |
| GB | 0.69 | 0.55 | 0.62 | 0.67 | 0.25 | 0.64 | 0.55 | 0.59 | 0.63 | 0.19 |

| Window 31 ||||||||||
|---|---|---|---|---|---|---|---|---|---|---|
| Model | Training Dataset ||||| Validation dataset |||||
| | Sens | Spec | Acc | AUROC | MCC | Sens | Spec | Acc | AUROC | MCC |
| RF | 0.68 | 0.55 | 0.62 | 0.67 | 0.23 | 0.64 | 0.53 | 0.59 | 0.63 | 0.17 |
| LGBM | 0.65 | 0.57 | 0.61 | 0.66 | 0.22 | 0.60 | 0.56 | 0.58 | 0.62 | 0.16 |
| NB | 0.64 | 0.55 | 0.60 | 0.62 | 0.19 | 0.58 | 0.57 | 0.58 | 0.59 | 0.15 |
| LR | 0.61 | 0.59 | 0.60 | 0.64 | 0.20 | 0.57 | 0.55 | 0.56 | 0.60 | 0.12 |
| GB | 0.69 | 0.55 | 0.62 | 0.66 | 0.23 | 0.67 | 0.52 | 0.60 | 0.63 | 0.19 |

# Supplementary Table S5: The complete result on different windows using PSSM profile + RSA + Binary profiles

LR, Logistic Regression; RF, Random Forest; NB, Naive Bayes; LGBM, Light Gradient Boosting Machine; GB, Gradient Boost; Sens, Sensitivity; Spec, Specificity; Acc, Accuracy; AUROC, Area Under the Receiver Operating Characteristic; MCC, Matthews Correlation Coefficient

## Window 7

| Model | Training Dataset | | | | | Validation dataset | | | | |
|---|---|---|---|---|---|---|---|---|---|---|
| | Sens | Spec | Acc | AUROC | MCC | Sens | Spec | Acc | AUROC | MCC |
| RF | 0.67 | 0.56 | 0.62 | 0.67 | 0.23 | 0.66 | 0.55 | 0.60 | 0.64 | 0.21 |
| LGBM | 0.67 | 0.57 | 0.62 | 0.66 | 0.25 | 0.65 | 0.54 | 0.60 | 0.63 | 0.19 |
| NB | 0.70 | 0.53 | 0.61 | 0.65 | 0.23 | 0.64 | 0.55 | 0.60 | 0.63 | 0.19 |
| GB | 0.71 | 0.55 | 0.63 | 0.67 | 0.26 | 0.68 | 0.51 | 0.59 | 0.64 | 0.19 |
| LR | 0.65 | 0.59 | 0.62 | 0.66 | 0.24 | 0.63 | 0.58 | 0.60 | 0.64 | 0.21 |

## Window 9

| Model | Training Dataset | | | | | Validation dataset | | | | |
|---|---|---|---|---|---|---|---|---|---|---|
| | Sens | Spec | Acc | AUROC | MCC | Sens | Spec | Acc | AUROC | MCC |
| RF | 0.69 | 0.56 | 0.63 | 0.67 | 0.26 | 0.64 | 0.55 | 0.60 | 0.65 | 0.20 |
| LGBM | 0.67 | 0.57 | 0.62 | 0.67 | 0.25 | 0.62 | 0.55 | 0.59 | 0.63 | 0.18 |
| NB | 0.67 | 0.55 | 0.61 | 0.65 | 0.22 | 0.63 | 0.56 | 0.59 | 0.63 | 0.19 |
| GB | 0.71 | 0.54 | 0.63 | 0.67 | 0.25 | 0.69 | 0.52 | 0.61 | 0.64 | 0.22 |
| LR | 0.64 | 0.59 | 0.61 | 0.66 | 0.22 | 0.61 | 0.58 | 0.60 | 0.64 | 0.20 |

## Window 11

| Model | Training Dataset | | | | | Validation dataset | | | | |
|---|---|---|---|---|---|---|---|---|---|---|
| | Sens | Spec | Acc | AUROC | MCC | Sens | Spec | Acc | AUROC | MCC |
| RF | 0.68 | 0.56 | 0.62 | 0.68 | 0.25 | 0.65 | 0.55 | 0.60 | 0.64 | 0.19 |
| LGBM | 0.67 | 0.58 | 0.62 | 0.67 | 0.25 | 0.63 | 0.56 | 0.60 | 0.63 | 0.19 |
| NB | 0.70 | 0.51 | 0.61 | 0.65 | 0.22 | 0.66 | 0.54 | 0.60 | 0.63 | 0.20 |
| LR | 0.63 | 0.59 | 0.61 | 0.66 | 0.22 | 0.61 | 0.58 | 0.59 | 0.64 | 0.19 |
| GB | 0.70 | 0.55 | 0.63 | 0.67 | 0.25 | 0.68 | 0.53 | 0.60 | 0.64 | 0.21 |

## Window 13

| Model | Training Dataset | | | | | Validation dataset | | | | |
|---|---|---|---|---|---|---|---|---|---|---|
| | Sens | Spec | Acc | AUROC | MCC | Sens | Spec | Acc | AUROC | MCC |
| RF | 0.68 | 0.57 | 0.62 | 0.67 | 0.25 | 0.65 | 0.56 | 0.60 | 0.64 | 0.21 |
| LGBM | 0.65 | 0.58 | 0.62 | 0.66 | 0.23 | 0.61 | 0.56 | 0.59 | 0.63 | 0.18 |
| NB | 0.70 | 0.50 | 0.60 | 0.64 | 0.21 | 0.65 | 0.53 | 0.59 | 0.62 | 0.18 |
| LR | 0.62 | 0.59 | 0.61 | 0.65 | 0.21 | 0.60 | 0.56 | 0.58 | 0.61 | 0.16 |
| GB | 0.70 | 0.55 | 0.63 | 0.67 | 0.26 | 0.66 | 0.52 | 0.59 | 0.63 | 0.18 |

| Window 15 | | | | | | | | | | |
|---|---|---|---|---|---|---|---|---|---|---|
| **Model** | **Training Dataset** | | | | | **Validation dataset** | | | | |
| | Sens | Spec | Acc | AUROC | MCC | Sens | Spec | Acc | AUROC | MCC |
| RF | 0.67 | 0.56 | 0.62 | 0.67 | 0.23 | 0.66 | 0.54 | 0.60 | 0.64 | 0.21 |
| LGBM | 0.66 | 0.58 | 0.62 | 0.66 | 0.24 | 0.63 | 0.55 | 0.59 | 0.63 | 0.18 |
| NB | 0.67 | 0.52 | 0.60 | 0.64 | 0.19 | 0.65 | 0.53 | 0.59 | 0.62 | 0.18 |
| LR | 0.62 | 0.59 | 0.61 | 0.65 | 0.21 | 0.62 | 0.55 | 0.58 | 0.62 | 0.17 |
| GB | 0.69 | 0.55 | 0.62 | 0.67 | 0.25 | 0.68 | 0.51 | 0.60 | 0.63 | 0.20 |

| Window 17 | | | | | | | | | | |
|---|---|---|---|---|---|---|---|---|---|---|
| **Model** | **Training Dataset** | | | | | **Validation dataset** | | | | |
| | Sens | Spec | Acc | AUROC | MCC | Sens | Spec | Acc | AUROC | MCC |
| RF | 0.67 | 0.57 | 0.62 | 0.67 | 0.24 | 0.63 | 0.55 | 0.59 | 0.63 | 0.18 |
| LGBM | 0.65 | 0.58 | 0.61 | 0.66 | 0.23 | 0.61 | 0.56 | 0.58 | 0.62 | 0.17 |
| NB | 0.68 | 0.52 | 0.60 | 0.64 | 0.21 | 0.62 | 0.54 | 0.58 | 0.60 | 0.16 |
| GB | 0.70 | 0.55 | 0.62 | 0.67 | 0.25 | 0.65 | 0.52 | 0.59 | 0.62 | 0.18 |
| LR | 0.61 | 0.59 | 0.60 | 0.65 | 0.20 | 0.58 | 0.54 | 0.56 | 0.60 | 0.12 |

| Window 19 | | | | | | | | | | |
|---|---|---|---|---|---|---|---|---|---|---|
| **Model** | **Training Dataset** | | | | | **Validation dataset** | | | | |
| | Sens | Spec | Acc | AUROC | MCC | Sens | Spec | Acc | AUROC | MCC |
| RF | 0.66 | 0.56 | 0.61 | 0.66 | 0.22 | 0.64 | 0.54 | 0.59 | 0.63 | 0.18 |
| LGBM | 0.65 | 0.56 | 0.61 | 0.65 | 0.21 | 0.61 | 0.56 | 0.59 | 0.63 | 0.18 |
| NB | 0.63 | 0.55 | 0.59 | 0.62 | 0.18 | 0.59 | 0.56 | 0.57 | 0.60 | 0.15 |
| GB | 0.69 | 0.55 | 0.62 | 0.66 | 0.24 | 0.65 | 0.51 | 0.58 | 0.62 | 0.16 |
| LR | 0.60 | 0.59 | 0.60 | 0.64 | 0.19 | 0.58 | 0.56 | 0.57 | 0.61 | 0.14 |

| Window 21 | | | | | | | | | | |
|---|---|---|---|---|---|---|---|---|---|---|
| **Model** | **Training Dataset** | | | | | **Validation dataset** | | | | |
| | Sens | Spec | Acc | AUROC | MCC | Sens | Spec | Acc | AUROC | MCC |
| RF | 0.68 | 0.57 | 0.62 | 0.67 | 0.25 | 0.64 | 0.54 | 0.59 | 0.64 | 0.18 |
| LGBM | 0.66 | 0.58 | 0.62 | 0.67 | 0.24 | 0.61 | 0.55 | 0.58 | 0.62 | 0.16 |
| NB | 0.67 | 0.52 | 0.59 | 0.63 | 0.19 | 0.59 | 0.54 | 0.57 | 0.60 | 0.13 |
| GB | 0.70 | 0.55 | 0.62 | 0.67 | 0.25 | 0.66 | 0.52 | 0.59 | 0.63 | 0.19 |
| LR | 0.61 | 0.59 | 0.60 | 0.65 | 0.20 | 0.59 | 0.55 | 0.57 | 0.60 | 0.14 |

| Window 23 | | | | | | | | | | |
|---|---|---|---|---|---|---|---|---|---|---|
| **Model** | **Training Dataset** | | | | | **Validation dataset** | | | | |
| | Sens | Spec | Acc | AUROC | MCC | Sens | Spec | Acc | AUROC | MCC |
| RF | 0.68 | 0.55 | 0.62 | 0.66 | 0.24 | 0.66 | 0.53 | 0.59 | 0.64 | 0.19 |
| LGBM | 0.67 | 0.58 | 0.63 | 0.67 | 0.25 | 0.61 | 0.55 | 0.58 | 0.62 | 0.15 |
| NB | 0.64 | 0.55 | 0.59 | 0.63 | 0.19 | 0.58 | 0.55 | 0.57 | 0.60 | 0.13 |

| Model | | | | | | | | | | |
|---|---|---|---|---|---|---|---|---|---|---|
| GB | 0.70 | 0.55 | 0.62 | 0.67 | 0.25 | 0.66 | 0.53 | 0.60 | 0.63 | 0.19 |
| LR | 0.60 | 0.58 | 0.59 | 0.64 | 0.19 | 0.57 | 0.55 | 0.56 | 0.60 | 0.13 |
| | | | | | | | | | | |
| **Window 25** | | | | | | | | | | |
| **Model** | **Training Dataset** | | | | | **Validation dataset** | | | | |
| | **Sens** | **Spec** | **Acc** | **AUROC** | **MCC** | **Sens** | **Spec** | **Acc** | **AUROC** | **MCC** |
| RF | 0.68 | 0.55 | 0.61 | 0.66 | 0.23 | 0.66 | 0.53 | 0.59 | 0.64 | 0.19 |
| LGBM | 0.65 | 0.56 | 0.61 | 0.65 | 0.21 | 0.60 | 0.56 | 0.58 | 0.62 | 0.16 |
| NB | 0.61 | 0.56 | 0.58 | 0.62 | 0.17 | 0.55 | 0.56 | 0.56 | 0.58 | 0.11 |
| LR | 0.60 | 0.57 | 0.58 | 0.62 | 0.17 | 0.58 | 0.54 | 0.56 | 0.59 | 0.12 |
| GB | 0.69 | 0.54 | 0.62 | 0.66 | 0.23 | 0.65 | 0.52 | 0.58 | 0.63 | 0.17 |
| | | | | | | | | | | |
| **Window 27** | | | | | | | | | | |
| **Model** | **Training Dataset** | | | | | **Validation dataset** | | | | |
| | **Sens** | **Spec** | **Acc** | **AUROC** | **MCC** | **Sens** | **Spec** | **Acc** | **AUROC** | **MCC** |
| RF | 0.66 | 0.56 | 0.61 | 0.66 | 0.22 | 0.65 | 0.54 | 0.59 | 0.64 | 0.18 |
| LGBM | 0.65 | 0.58 | 0.61 | 0.66 | 0.23 | 0.61 | 0.55 | 0.58 | 0.62 | 0.16 |
| NB | 0.61 | 0.55 | 0.58 | 0.61 | 0.16 | 0.56 | 0.56 | 0.56 | 0.59 | 0.13 |
| LR | 0.59 | 0.58 | 0.59 | 0.63 | 0.17 | 0.58 | 0.56 | 0.57 | 0.60 | 0.14 |
| GB | 0.68 | 0.54 | 0.61 | 0.66 | 0.23 | 0.67 | 0.53 | 0.60 | 0.63 | 0.20 |
| | | | | | | | | | | |
| **Window 29** | | | | | | | | | | |
| **Model** | **Training Dataset** | | | | | **Validation dataset** | | | | |
| | **Sens** | **Spec** | **Acc** | **AUROC** | **MCC** | **Sens** | **Spec** | **Acc** | **AUROC** | **MCC** |
| RF | 0.66 | 0.55 | 0.61 | 0.66 | 0.22 | 0.62 | 0.55 | 0.59 | 0.63 | 0.17 |
| LGBM | 0.66 | 0.58 | 0.62 | 0.66 | 0.24 | 0.61 | 0.57 | 0.59 | 0.62 | 0.18 |
| NB | 0.62 | 0.54 | 0.58 | 0.61 | 0.17 | 0.58 | 0.55 | 0.57 | 0.59 | 0.13 |
| LR | 0.60 | 0.58 | 0.59 | 0.63 | 0.18 | 0.59 | 0.54 | 0.57 | 0.59 | 0.14 |
| GB | 0.68 | 0.55 | 0.61 | 0.66 | 0.23 | 0.65 | 0.53 | 0.59 | 0.63 | 0.18 |
| | | | | | | | | | | |
| **Window 31** | | | | | | | | | | |
| **Model** | **Training Dataset** | | | | | **Validation dataset** | | | | |
| | **Sens** | **Spec** | **Acc** | **AUROC** | **MCC** | **Sens** | **Spec** | **Acc** | **AUROC** | **MCC** |
| RF | 0.67 | 0.55 | 0.61 | 0.66 | 0.22 | 0.64 | 0.54 | 0.59 | 0.63 | 0.18 |
| LGBM | 0.65 | 0.57 | 0.61 | 0.65 | 0.22 | 0.63 | 0.55 | 0.59 | 0.63 | 0.18 |
| NB | 0.60 | 0.57 | 0.59 | 0.62 | 0.18 | 0.57 | 0.57 | 0.57 | 0.60 | 0.14 |
| LR | 0.60 | 0.58 | 0.59 | 0.62 | 0.18 | 0.59 | 0.55 | 0.57 | 0.59 | 0.13 |
| GB | 0.69 | 0.54 | 0.61 | 0.66 | 0.23 | 0.67 | 0.52 | 0.59 | 0.63 | 0.18 |